\documentclass[universe,article,accept,pdftex,moreauthors]{Definitions/mdpi} 
\firstpage{1} 
\pubvolume{1}
\issuenum{1}
\articlenumber{0}
\pubyear{2022}
\copyrightyear{2022}
\externaleditor{Academic Editor: }
\datereceived{} 
\dateaccepted{} 
\datepublished{} 
\hreflink{https://doi.org/} 

\Title{Exploring Anisotropic Lorentz Invariance Violation from the Spectral-Lag Transitions of Gamma-Ray Bursts}

\TitleCitation{Exploring Anisotropic Lorentz Invariance Violation from the Spectral-Lag Transitions of Gamma-Ray Bursts}

\newcommand{\prl}{Phys. Rev. Lett.}
\newcommand{\prd}{Phys. Rev. D}
\newcommand{\apjl}{Astrophys. J. Lett.}
\newcommand{\apj}{Astrophys. J.}
\newcommand{\mnras}{Mon. Not. R. Astron. Soc.}
\newcommand{\aap}{Astron. Astrophys.}
\newcommand{\nat}{Nature}
\newcommand{\jcap}{J. Cosmol. Astropart. Phys.}
\newcommand{\pasp}{Publ. Astron. Soc. Pac.}

\Author{Jin-Nan Wei $^{1,2,3}$, Zi-Ke Liu $^{4,5}$, Jun-Jie Wei $^{1,2,3,}$*\orcidA{}, Bin-Bin Zhang $^{4,5}$\orcidB{} and Xue-Feng Wu $^{1,2}$\orcidC{}}

\AuthorNames{Jin-Nan Wei, Zi-Ke Liu, Jun-Jie Wei, Bin-Bin Zhang and Xue-Feng Wu}

\AuthorCitation{Wei, J.-N.; Liu, Z.-K.; Wei, J.-J.; Zhang, B.-B.; Wu, X.-F.}

\address{%
$^{1}$ \quad Purple Mountain Observatory, Chinese Academy of Sciences, Nanjing 210023, China;\linebreak weijn@pmo.ac.cn (J.-N.W.); xfwu@pmo.ac.cn (X.-F.W.)\\
$^{2}$ \quad School of Astronomy and Space Sciences, University of Science and Technology of China, Hefei 230026, China\\
$^{3}$ \quad Guangxi Key Laboratory for Relativistic Astrophysics, Nanning 530004, China\\
$^{4}$ \quad School of Astronomy and Space Science, Nanjing University, Nanjing 210023, China;\linebreak zkliu@smail.nju.edu.cn (Z.-K.L.); bbzhang@nju.edu.cn (B.-B.Z.)\\
$^{5}$ \quad Key Laboratory of Modern Astronomy and Astrophysics (Nanjing University), Ministry of Education, Nanjing 210023, China}

\corres{Correspondence: jjwei@pmo.ac.cn}

\abstract{The observed spectral lags of gamma-ray bursts (GRBs) have been widely used to explore 
possible violations of Lorentz invariance. However, these studies were generally performed by 
concentrating on the rough time lag of a single highest-energy photon and ignoring the intrinsic 
time lag at the source. A new way to test nonbirefringent Lorentz-violating effects has been 
proposed by analyzing the multi-photon spectral-lag behavior of a GRB that displays a 
positive-to-negative transition. This method gives both a plausible description of the intrinsic 
energy-dependent time lag and comparatively robust constraints on Lorentz-violating effects. 
In this work, we conduct a systematic search for Lorentz-violating photon dispersion from the
spectral-lag transition features of 32 GRBs. By fitting the spectral-lag data of these 32 GRBs,
we place constraints on a variety of isotropic and anisotropic Lorentz-violating coefficients with
mass dimension $d=6$ and $8$. While our dispersion constraints are not competitive with existing 
bounds, they have the promise to complement the full coefficient space.}


\begin{document}

\section{Introduction}

\textls[-15]{Lorentz invariance, the foundational symmetry of Einstein's relativity, has survived in a wide range of tests over 
the past century \citep{2011RvMP...83...11K}. However, many quantum gravity models seeking to unify 
general relativity and quantum theory predict that Lorentz symmetry may be violated
at energies approaching the Planck scale $E_{\rm Pl}=\sqrt{\hbar c^{5}/G}\simeq1.22\times10^{19}$ GeV \citep{1989PhRvD..39..683K,1991NuPhB.359..545K,1995PhRvD..51.3923K,2005LRR.....8....5M,2013LRR....16....5A,2014RPPh...77f2901T,2021FrPhy..1644300W,2022Univ....8..323H}. 
While these energies are unreachable experimentally, tiny deviations from Lorentz invariance at attainable 
energies can accumulate to detectable levels over sufficiently large distances. Astrophysical observations 
involving long propagation distances can therefore provide precision tests of Lorentz invariance.}

In the photon sector, one effect of Lorentz invariance violation (LIV) is an energy-dependent vacuum dispersion
of light, which causes arrival-time delays of photons with different energies originating from a given astrophysical
source \citep{1998Natur.393..763A,2000ApJ...535..139E,2005PhLB..625...13P,2006APh....25..402E,2008JCAP...01..031J,2008ApJ...689L...1K,2009PhRvD..80a5020K,2009Sci...323.1688A,2013APh....43...50E,2013PhRvD..87l2001V,2015PhRvD..92d5016K,2017ApJ...834L..13W,2017ApJ...842..115W,2017ApJ...851..127W,2021ApJ...906....8D,2022arXiv220209999L,2022ApJ...924L..29X}. 
LIV models can also lead to vacuum birefringence, which produces an energy-dependent rotation of the polarization 
vector of linearly polarized light \citep{2001PhRvL..87y1304K,2006PhRvL..97n0401K,2007PhRvL..99a1601K,2013PhRvL.110t1601K,2009JCAP...08..021G,2011PhRvD..83l1301L,2011APh....35...95S,2012PhRvL.109x1104T,2017PhRvD..95h3013K,2019PhRvD..99c5045F,2019MNRAS.485.2401W,2021Galax...9...44Z}. Generally, these effects can be anisotropic, such that arrival-time 
differences and polarization rotations possess a direction dependence and require observations of point sources 
along many lines of sight or measurements of extended sources such as the cosmic microwave background to fully explore 
the LIV model parameter space \citep{2008ApJ...689L...1K,2009PhRvD..80a5020K,2019PhRvD..99c5045F}.

The Standard-Model Extension (SME) is an effective field theory that characterizes Lorentz and CPT violations at
attainable energies \citep{1997PhRvD..55.6760C,1998PhRvD..58k6002C,2004PhRvD..69j5009K}. It considers additional
Lorentz and/or CPT-violating terms to the SME Lagrange density, which can be ordered by the mass dimension $d$
of the tensor operator \citep{2009PhRvD..80a5020K}. Photon vacuum dispersion introduced by operators of dimension 
$d$ ($\neq 4$) is proportional to $(E/E_{\rm Pl})^{d-4}$. Lorentz-violating operators with even $d$ preserve CPT
symmetry, while those with odd $d$ break CPT. All $(d-1)^{2}$ coefficients of odd $d$ produce both vacuum 
dispersion and birefringence, whereas for each even $d$ there is a subset of $(d-1)^{2}$ nonbirefringent but 
dispersive Lorentz-violating coefficients. The latter can be well constrained through vacuum dispersion 
time-of-flight measurements. It should be stressed that at least $(d-1)^{2}$ sources distributed 
evenly in the sky are needed to fully constrain  the anisotropic Lorentz-violating coefficient space 
for a given $d$. 
In contrast, only one source is required to fully constrain the corresponding coefficient 
in the isotropic LIV limit. That is, the restriction to isotropic LIV disregards $d(d-2)$ possible effects 
from anisotropic violations at each $d$.

Thanks to their small variability time scales, large cosmological distances, and very high-energy photons, 
gamma-ray bursts (GRBs) are viewed as one of the ideal probes for testing Lorentz invariance through the dispersion
method \citep{1998Natur.393..763A,2006APh....25..402E,2008JCAP...01..031J}. Direction-dependent limits on several
combinations of coefficients for Lorentz violation have been placed using vacuum-dispersion constraints from GRBs.
For example, limits on combinations of the 25 $d=6$ nonbirefringent Lorentz-violating coefficients have been derived
by studying the dispersion of light in observations of GRB 021206 \citep{2004ApJ...611L..77B,2008ApJ...689L...1K},
GRB 080916C \citep{2009Sci...323.1688A,2009PhRvD..80a5020K}, four bright GRBs \citep{2013PhRvD..87l2001V}, 
GRB 160625B \citep{2017ApJ...842..115W}, and GRB 190114C \citep{2021ApJ...906....8D}. Bounds on combinations of 
the 49 $d=8$ coefficients for nonbirefringent vacuum dispersion have been derived from GRB 021206 
\citep{2004ApJ...611L..77B,2008ApJ...689L...1K}, GRB 080916C \citep{2009Sci...323.1688A,2009PhRvD..80a5020K}, 
GRB 160625B \citep{2017ApJ...842..115W}, and GRB 190114C \citep{2021ApJ...906....8D}. However, most of these
studies limit attention to the time delay induced by nonbirefringent Lorentz-violating effects, while ignoring 
possible source-intrinsic time delays. Furthermore, the limits from GRBs are based on the rough time delay of 
a single highest-energy photon. To obtain reliable LIV limits, it is desirable to use the true time lags of 
high-quality and high-energy light curves in different energy multi-photon bands. 

\textls[-15]{In two previous papers \citep{2017ApJ...842..115W,2021ApJ...906....8D}, we derived new direction-dependent limits
on combinations of coefficients for Lorentz-violating vacuum dispersion using the peculiar time-of-flight measurements
of GRB 160625B and GRB 190114C, which both have obvious transitions from positive to negative spectral lags.
Spectral lag, which is defined as the arrival-time difference of high- and low-energy photons, is a ubiquitous 
feature in GRBs \citep{1997ApJ...486..928B,1996ApJ...459..393N,1995A&A...300..746C}. Conventionally, the spectral lag 
is considered to be positive when high-energy photons arrive earlier than the low-energy ones. 
By fitting the 
spectral-lag behaviors of GRB 160625B and GRB 190114C, we obtained both a reasonable formulation of the intrinsic
energy-dependent time lag and robust constraints on a variety of isotropic and anisotropic Lorentz-violating 
coefficients with mass dimension $d=6$ and $8$ \citep{2017ApJ...842..115W,2021ApJ...906....8D}. In this work,
we analyze the spectral-lag transition features of 32 $Fermi$ GRBs \citep{2022arXiv220209999L}, and derive limits 
on photon vacuum dispersion for all of them. We combine these limits with previous results in order to fully
constrain the nonbirefringent Lorentz-violating coefficients with $d=6$ and $8$.} 

The paper is structured as follows. In Section~\ref{sec:framework}, we briefly describe the theoretical framework 
of vacuum dispersion in the SME. In Section~\ref{sec:LIV}, we introduce our analysis method, and then present 
our resulting constraints on the Lorentz-violating coefficients. The physical implications of our
results are discussed in Section~\ref{sec:Discussion}. Finally, we summarize our main conclusions
in Section~\ref{sec:summary}. 

\section{Theoretical Framework}
\label{sec:framework}

In the SME framework, the LIV-induced modifications to the photon dispersion relation can be described in the form
\citep{2008ApJ...689L...1K,2009PhRvD..80a5020K}
\begin{equation}
  E(p) \simeq
\bigl(1 - \varsigma^0
\pm \sqrt{(\varsigma^1)^2 + (\varsigma^2)^2 + (\varsigma^3)^2}\bigr)
\, p\;,
\label{mdr}
\end{equation}
where $p$ is the photon momentum. The symbols $\varsigma^{0}, \varsigma^{1}, \varsigma^{2}$, and $\varsigma^{3}$ are the combinations of coefficients for LIV that depend on the momentum and direction of propagation. These four combinations
can be decomposed on a spherical harmonic basis to yield 
\begin{equation}
\begin{aligned}
  \varsigma^0 &=
\sum_{djm}p^{d-4} {}_{0}Y_{jm}(\hat{\textbf{\emph{n}}})c_{(I)jm}^{(d)},
\\
  \varsigma^1 \pm i\varsigma^2 &=
\sum_{djm}p^{d-4} {}_{\mp2}{Y}_{jm}(\hat{\textbf{\emph{n}}})
\left(k_{(E)jm}^{(d)} \mp ik_{(B)jm}^{(d)}\right),
\\
  \varsigma^3 &=
\sum_{djm}p^{d-4} {}_{0}Y_{jm}(\hat{\textbf{\emph{n}}})k_{(V)jm}^{(d)},
\label{eq:sigma0}
\end{aligned}
\end{equation}
where $\hat{\textbf{\emph{n}}}$ is the direction of the source and ${}_{s}{Y}_{jm}(\hat{\textbf{\emph{n}}})$
represents spin-weighted harmonics of spin weight $s$. The coordinates $(\theta,\;\phi)$ of $\hat{\textbf{\emph{n}}}$
are in a Sun-centered celestial-equatorial frame~\citep{2002PhRvD..66e6005K}, such that $\theta = (90^{\circ}-\rm{Dec.})$
and $\phi =$ R.A., where R.A.\ and Dec.\ are the right ascension and declination of the astrophysical source, respectively.

With the above decomposition, all types of LIV for vacuum propagation can be characterized using four sets of spherical
coefficients: $c_{(I)jm}^{(d)}$, $k_{(E)jm}^{(d)}$, and $k_{(B)jm}^{(d)}$ for CPT-even effects and $k_{(V)jm}^{(d)}$
for CPT-odd effects. For each coefficient, the relevant Lorentz-violating operator has mass dimension $d$ and eigenvalues
of total angular momentum written as $jm$. The coefficients $c_{(I)jm}^{(d)}$ are associated with CPT-even operators
causing dispersion without leading-order birefringence, while nonzero coefficients $k_{(E)jm}^{(d)}$, $k_{(B)jm}^{(d)}$, and
$k_{(V)jm}^{(d)}$ produce birefringence. In the present work, we focus on the nonbirefringent vacuum dispersion coefficients
$c_{(I)jm}^{(d)}$. Setting all other coefficients for birefringent propagation to zero, the group-velocity defect including
anisotropies is given by
\begin{equation}
\delta v_g = - \sum_{djm} (d-3)
E^{d-4} \, {}_{0}Y_{jm}(\hat{\textbf{\emph{n}}}) c_{(I)jm}^{(d)}\;,
\end{equation}
in terms of the photon energy $E$. Note that the factor $(d-3)$ refers to the difference between group and phase velocities
\citep{2017ApJ...842..115W}. For even $d\geq6$, nonzero values of $c_{(I)jm}^{(d)}$ imply an energy-dependent vacuum dispersion
of light, so two photons with different energies ($E_{\rm h}>E_{\rm l}$) emitted simultaneously
from the same astrophysical source at redshift $z$ would be observed at different times.
The arrival-time difference can therefore be derived as \citep{2008ApJ...689L...1K}
\begin{equation}
\begin{aligned}
\bigtriangleup{t_{\rm LIV}} &= t_{\rm l} - t_{\rm h}\\ &\approx -(d - 3)\left(E_{\rm h}^{d-4} - E_{\rm l}^{d-4}\right)
 \int^{z}_{0} \frac{(1 + z')^{d - 4}}{H(z')}{\rm d} z'
\sum_{jm} {}_{0}Y_{jm}(\hat{\textbf{\emph{n}}}) c_{(I)jm}^{(d)} \;,
\label{eq:tLIV}
\end{aligned}
\end{equation}
where $t_{\rm l}$ and $t_{\rm h}$ are the arrival times of photons with observed energies $E_{\rm l}$ and $E_{\rm h}$,
respectively. In the flat $\Lambda$CDM model, the Hubble expansion rate $H(z)$ is expressed as
$H(z) = H_0\left[\Omega_m(1+z)^3 + \Omega_\Lambda\right]^{1/2}$, where $H_{0}=67.36$ km $\rm s^{-1}$ $\rm Mpc^{-1}$
is the Hubble constant, $\Omega_{m}=0.315$ is the matter density, and $\Omega_{\Lambda}=1-\Omega_{m}$ is the
cosmological constant energy density \citep{2020A&A...641A...6P}.
In the SME case of a direction-dependent LIV, we constrain the combination $\sum_{jm} {}_{0}Y_{jm}(\hat{\textbf{\emph{n}}}) c_{(I)jm}^{(d)}$ as a whole. For the limiting case of
the vacuum isotropic model ($j=m=0$), all the terms in the combination become zero except
${}_{0}Y_{00}=Y_{00}=\sqrt{1/(4\pi)}$. In that case, we constrain a single $c_{(I)00}^{(d)}$ coefficient.

\section{Constraints on Anisotropic LIV}
\label{sec:LIV}
By systematically analyzing the spectral lags of 135 $Fermi$ long GRBs with redshift measurement,
Liu et al. \cite{2022arXiv220209999L} identified 32 of them having well-defined transitions from 
positive to negative spectral lags. For each GRB, Liu et al. \cite{2022arXiv220209999L} extracted its
multi-band light curves and calculated the spectral time lags for any pair of light curves between
the lowest-energy band and any other higher-energy bands. In Figure~\ref{fig1}, we plot the time lags of 
each burst as a function of energy. One can see that all GRBs exhibit a positive-to-negative lag 
transition. In this section, we utilize the spectral-lag transitions of 
these 32 GRBs to pose direction-dependent constraints on combinations of nonbirefringent
Lorentz-violating coefficients $c_{(I)jm}^{(6)}$ and $c_{(I)jm}^{(8)}$. The 32 GRBs used in our study 
are listed in Table~\ref{tab1}, which includes the following information for each burst:
its name, the right ascension coordinate, the declination coordinate, and the redshift $z$.

\begin{table}[H]
\caption{List of 32 GRBs with spectral-lag transitions.\label{tab1}}
\begin{tabularx}{\textwidth}{lCCCC}
\toprule
     &  \textbf{R.A.}  & \textbf{Dec.} & \textbf{Redshift} \\
 \textbf{Name} & \textbf{J2000 [}\boldmath{$^{\circ}$}\textbf{]}  & \textbf{J2000 [}\boldmath{$^{\circ}$}\textbf{]} & \boldmath{$z$} \textsuperscript{\textbf{1}} & \textbf{Ref.} \textsuperscript{\textbf{2}} \\
\midrule
GRB 080916C & 119.8 & $-$56.6 & 4.35 & \cite{von_Kienlin_2020} \\
GRB 081221 & 15.8 & $-$24.5 & 2.26 & \cite{von_Kienlin_2020} \\
GRB 090328 & 155.7 & +33.4 & 0.736 & \cite{von_Kienlin_2020} \\
GRB 090618 & 294.0 & +78.4 & 0.54 & \cite{von_Kienlin_2020} \\
GRB 090926A & 353.4 & $-$66.3 & 2.1062 & \cite{von_Kienlin_2020} \\
GRB 091003A & 251.5 & $-$36.6 & 0.8969 & \cite{von_Kienlin_2020} \\
GRB 100728A & 88.8 & $-$15.3 & 1.567 & \cite{von_Kienlin_2020} \\
GRB 120119A & 120.0 & $-$9.8 & 1.728 & \cite{von_Kienlin_2020} \\
GRB 130427A & 173.1 & +27.7 & 0.3399 & \cite{von_Kienlin_2020} \\
GRB 130518A & 355.7 & +47.5 & 2.488 & \cite{von_Kienlin_2020} \\
GRB 130925A & 41.2 & $-$26.1 & 0.347 & \cite{von_Kienlin_2020} \\
GRB 131108A & 156.5 & +9.7 & 2.40 & \cite{von_Kienlin_2020} \\
GRB 131231A & 10.6 & $-$1.6 & 0.642 & \cite{von_Kienlin_2020} \\
GRB 140206A & 145.3 & +66.8 & 2.73 & \cite{von_Kienlin_2020} \\
GRB 140508A & 255.5 & +46.8 & 1.027 & \cite{von_Kienlin_2020} \\
GRB 141028A & 322.6 & $-$0.2 & 2.33 & \cite{von_Kienlin_2020} \\
GRB 150314A & 126.7 & +63.8 & 1.758 & \cite{von_Kienlin_2020} \\
GRB 150403A & 311.5 & $-$62.7 & 2.06 & \cite{von_Kienlin_2020} \\
GRB 150514A & 74.8 & $-$60.9 & 0.807 & \cite{von_Kienlin_2020} \\
GRB 150821A & 341.9 & $-$57.9 & 0.755 & \cite{von_Kienlin_2020} \\
GRB 160509A & 310.1 & +76.0 & 1.17 & \cite{von_Kienlin_2020} \\
GRB 160625B & 308.6 & +6.9 & 1.41 & \cite{von_Kienlin_2020} \\
GRB 171010A & 66.6 & $-$10.5 & 0.3285 & \cite{von_Kienlin_2020} \\
GRB 180703A & 6.5 & $-$67.1 & 0.6678 & \cite{von_Kienlin_2020} \\
GRB 180720B & 0.59 & $-$3.0 & 0.654 & \cite{Ajello_2019} \\
GRB 190114C & 54.5 & $-$26.9 & 0.425 & \cite{2019GCN.23695....1S} \\
GRB 200613A & 153.0 & +45.8 & 1.22 & \cite{2020GCN.28000....1B} \\
GRB 200829A & 251.1 & +72.4 & 1.25 & \cite{2020GCN.28338....1O} \\
GRB 201216C & 16.4 & +16.5 & 1.10 & \cite{2020GCN.29077....1V} \\
GRB 210204A & 109.1 & +9.7 & 0.876 & \cite{2021GCN.29432....1X} \\
GRB 210610B & 243.9 & +14.4 & 1.13 & \cite{2021GCN.30194....1D} \\
GRB 210619B & 319.7 & +33.9 & 1.937 & \cite{2021GCN.30272....1D} \\
\bottomrule
\end{tabularx}

\noindent{\footnotesize{\textsuperscript{1} All source redshifts were obtained from Liu et al. 
\cite{2022arXiv220209999L} and references therein. \textsuperscript{2} Individual references are given for the coordinates.}} 
\end{table}
\vspace{-6pt}
\begin{figure}[H]
\centering
\subfigure{\includegraphics[width=4.2cm]{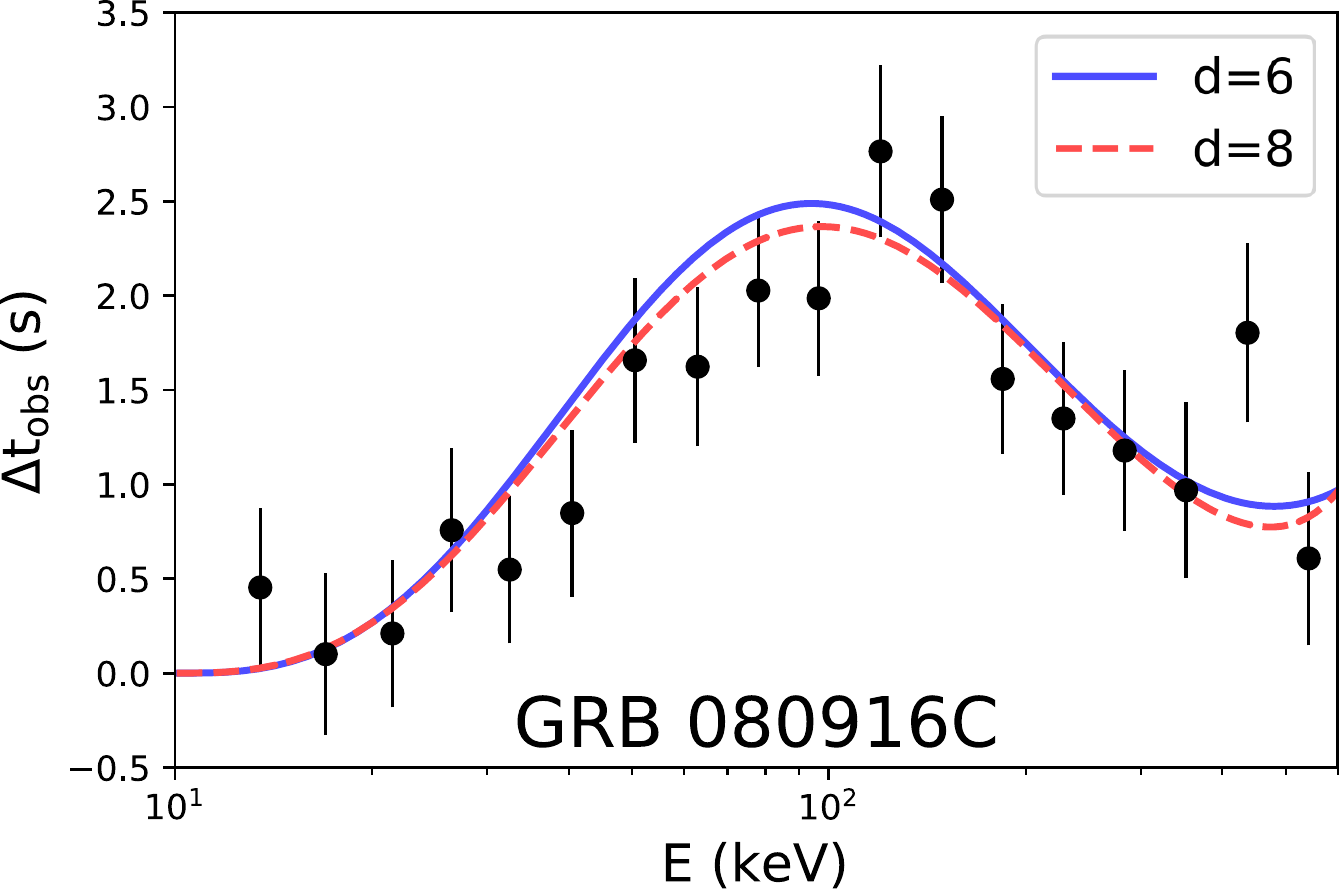}}
\subfigure{\includegraphics[width=4.2cm]{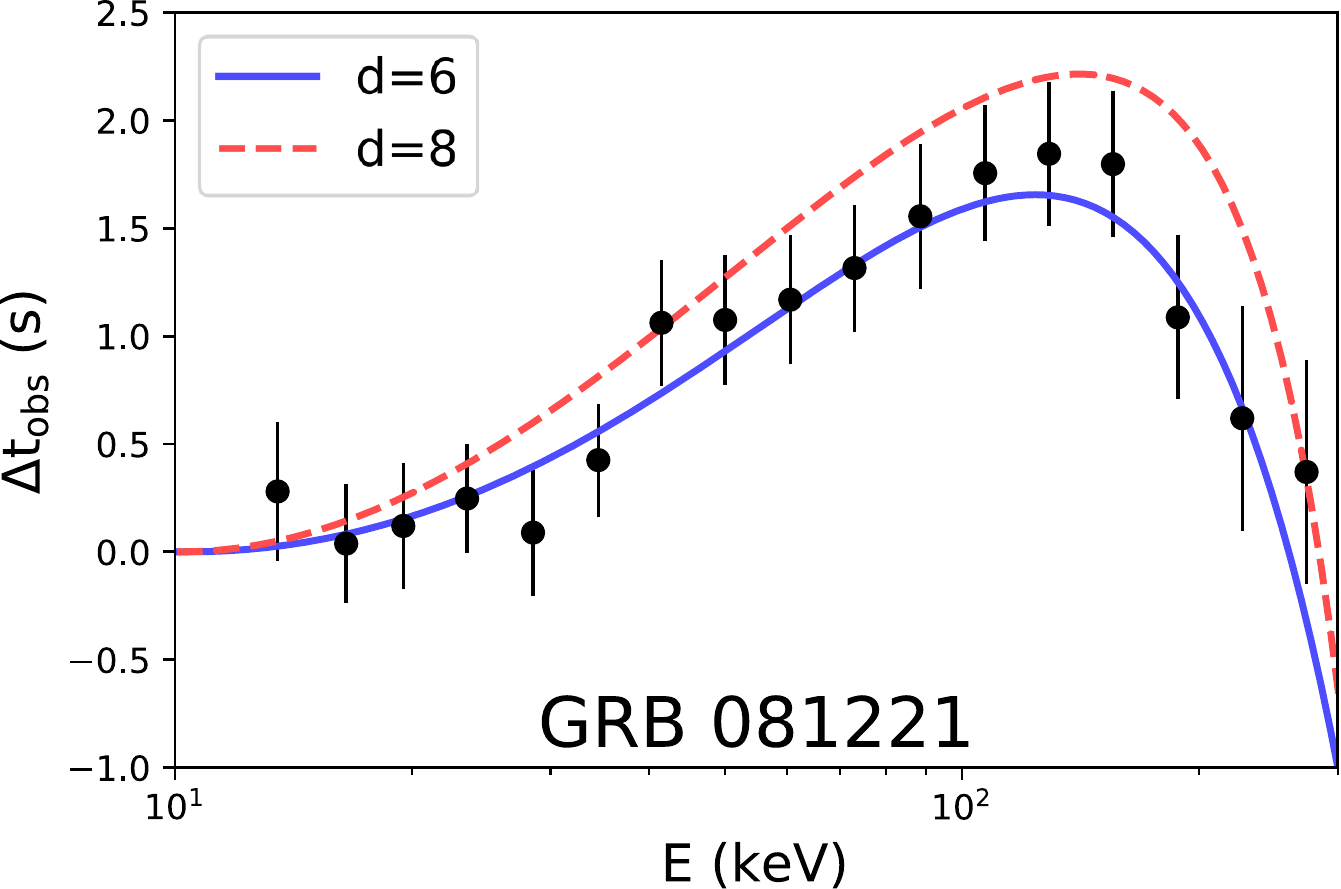}}
\subfigure{\includegraphics[width=4.2cm]{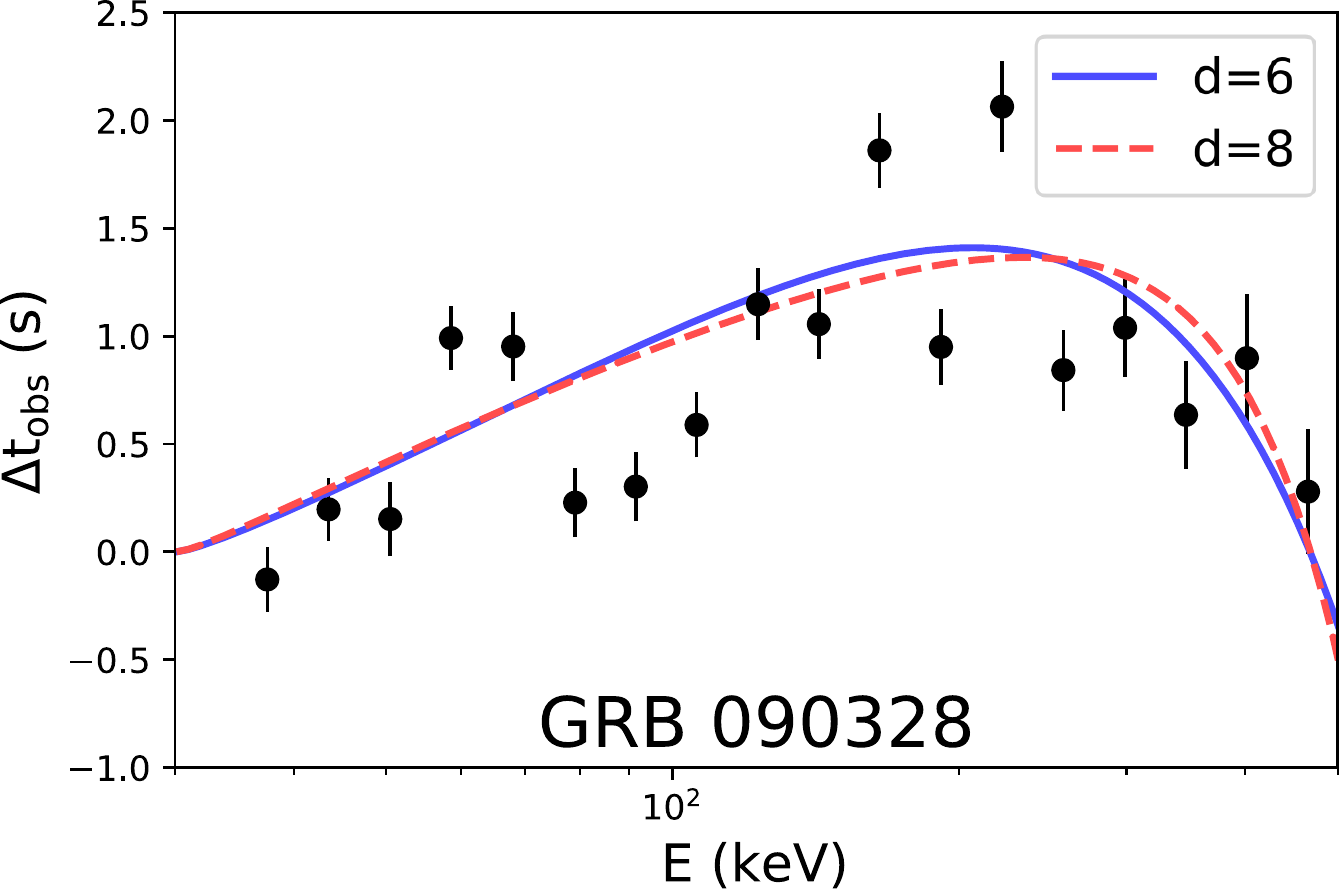}}
\subfigure{\includegraphics[width=4.2cm]{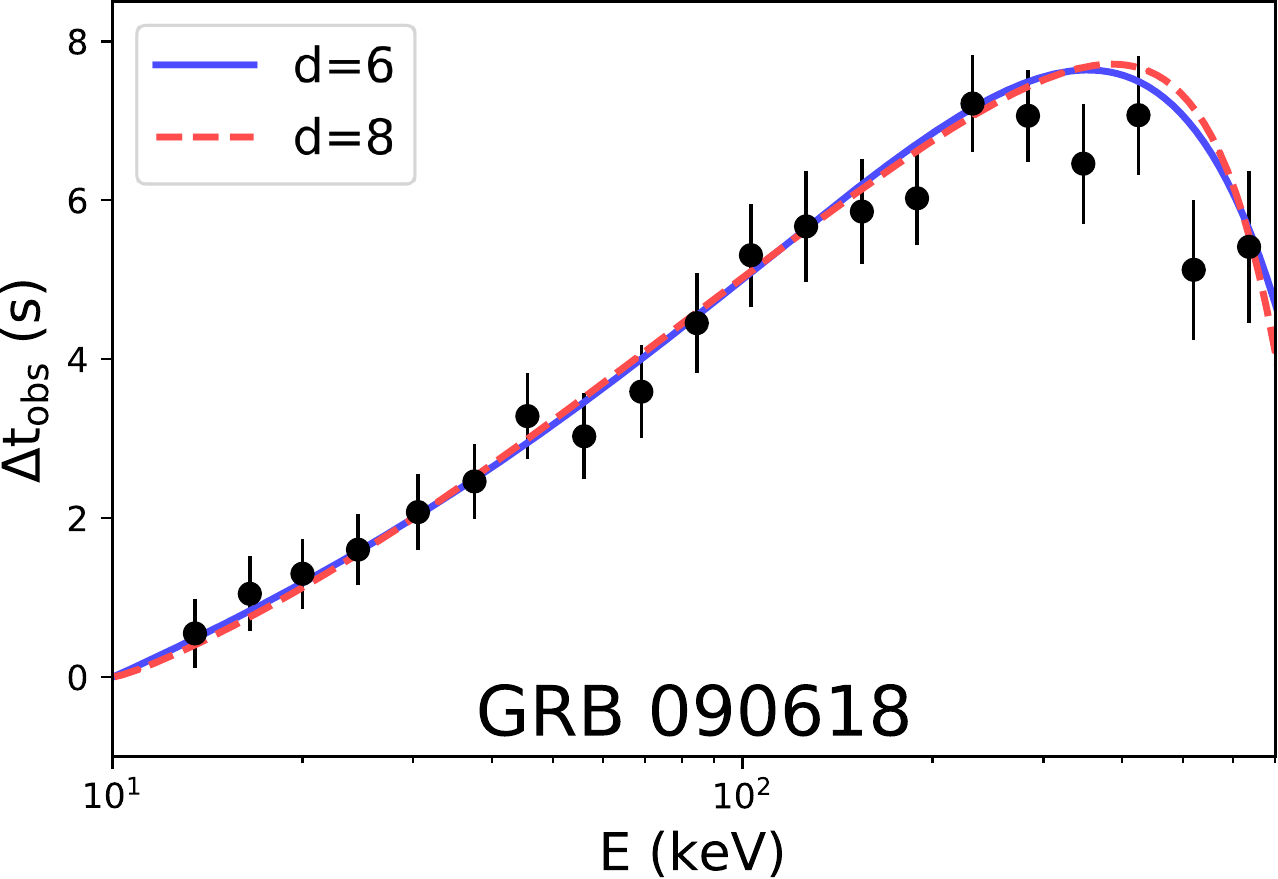}}
\subfigure{\includegraphics[width=4.2cm]{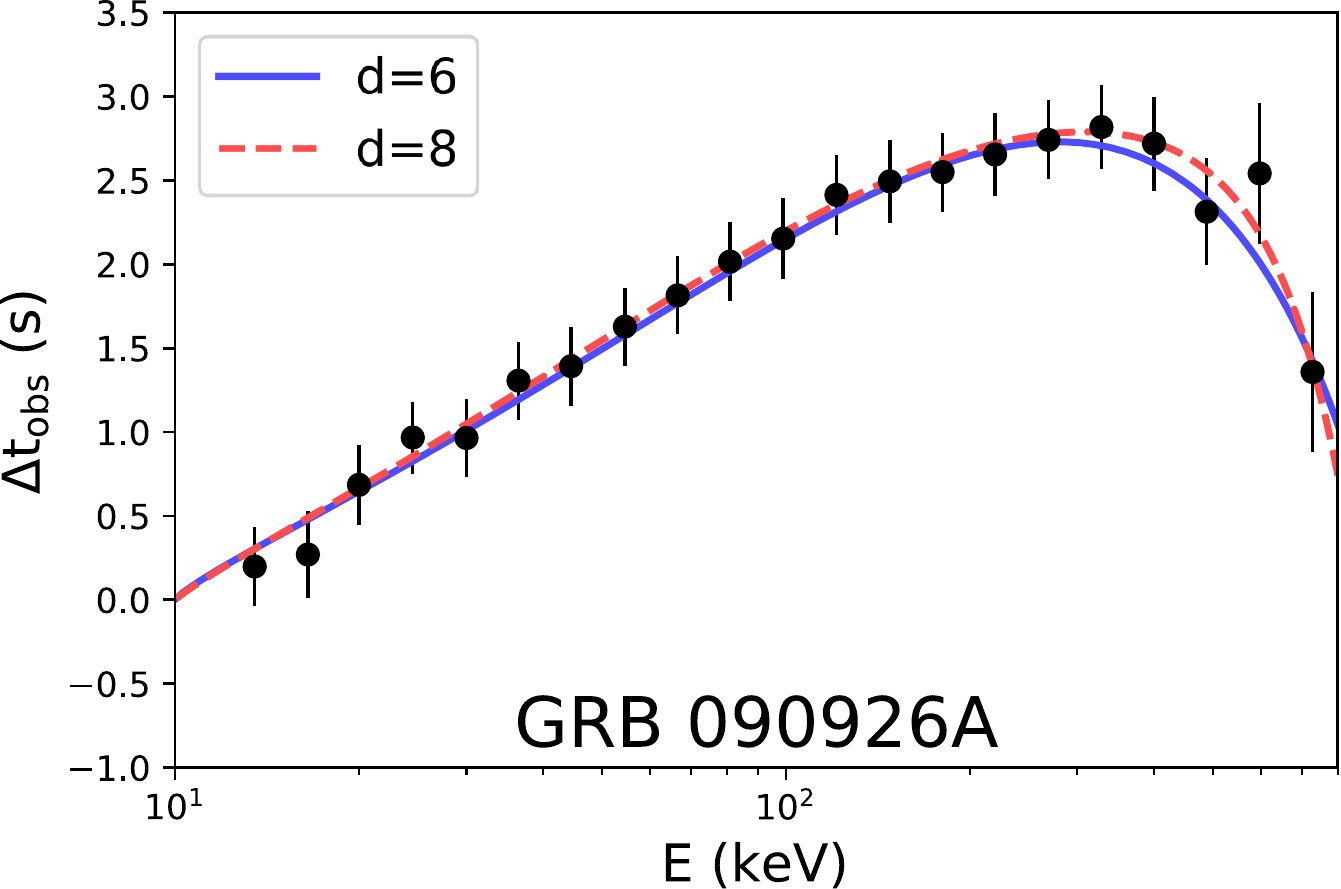}}
\subfigure{\includegraphics[width=4.2cm]{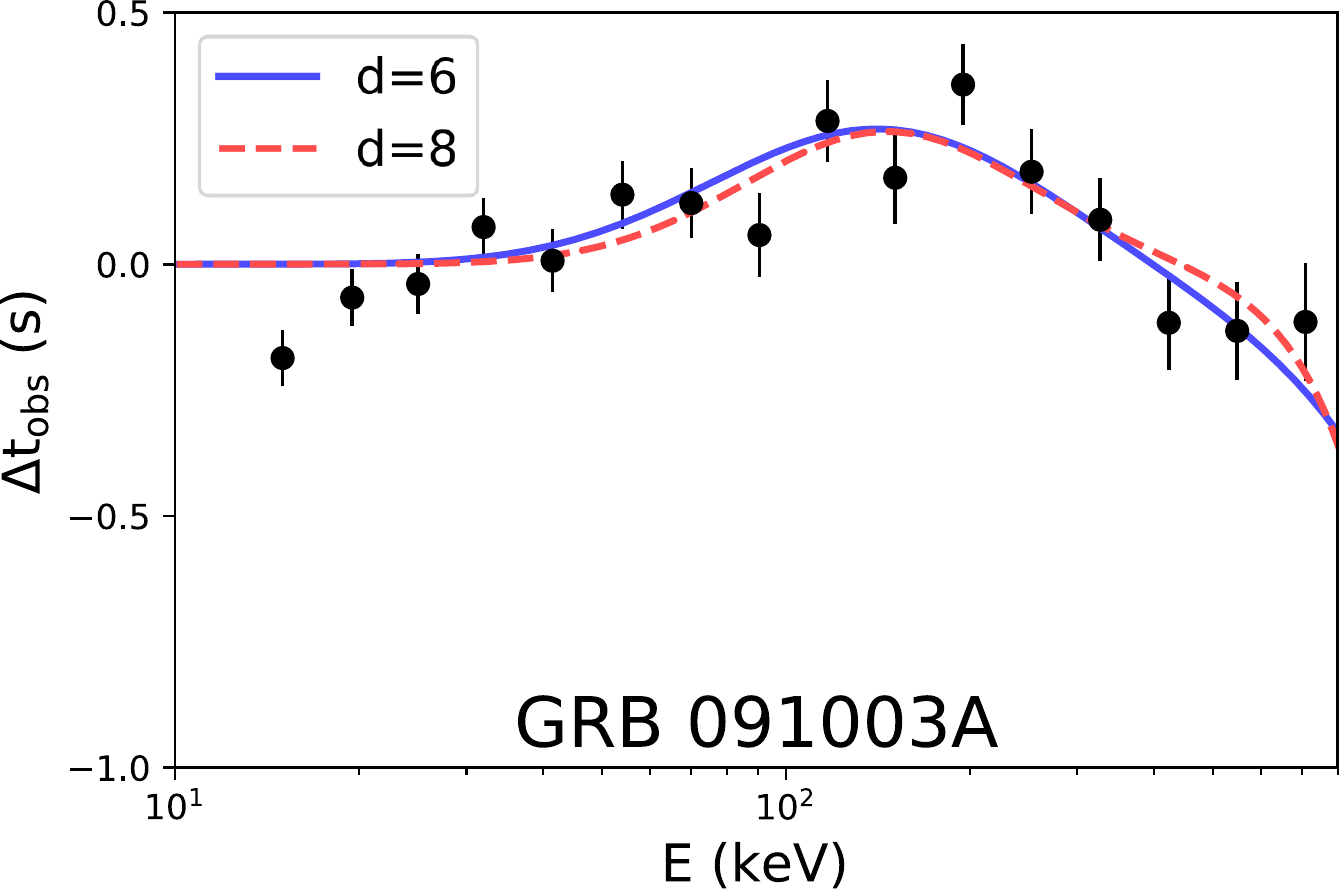}}
\subfigure{\includegraphics[width=4.2cm]{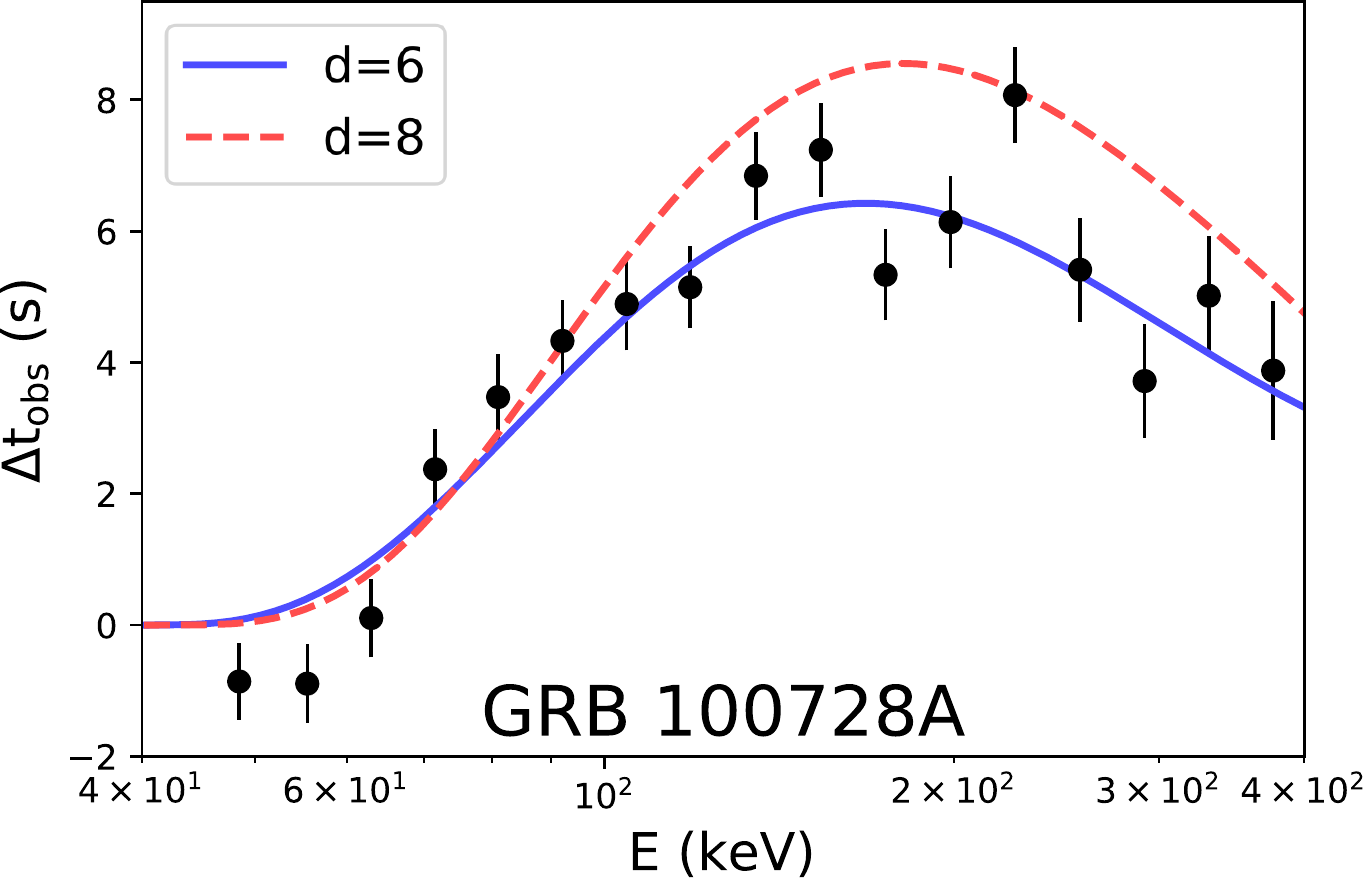}}
\subfigure{\includegraphics[width=4.2cm]{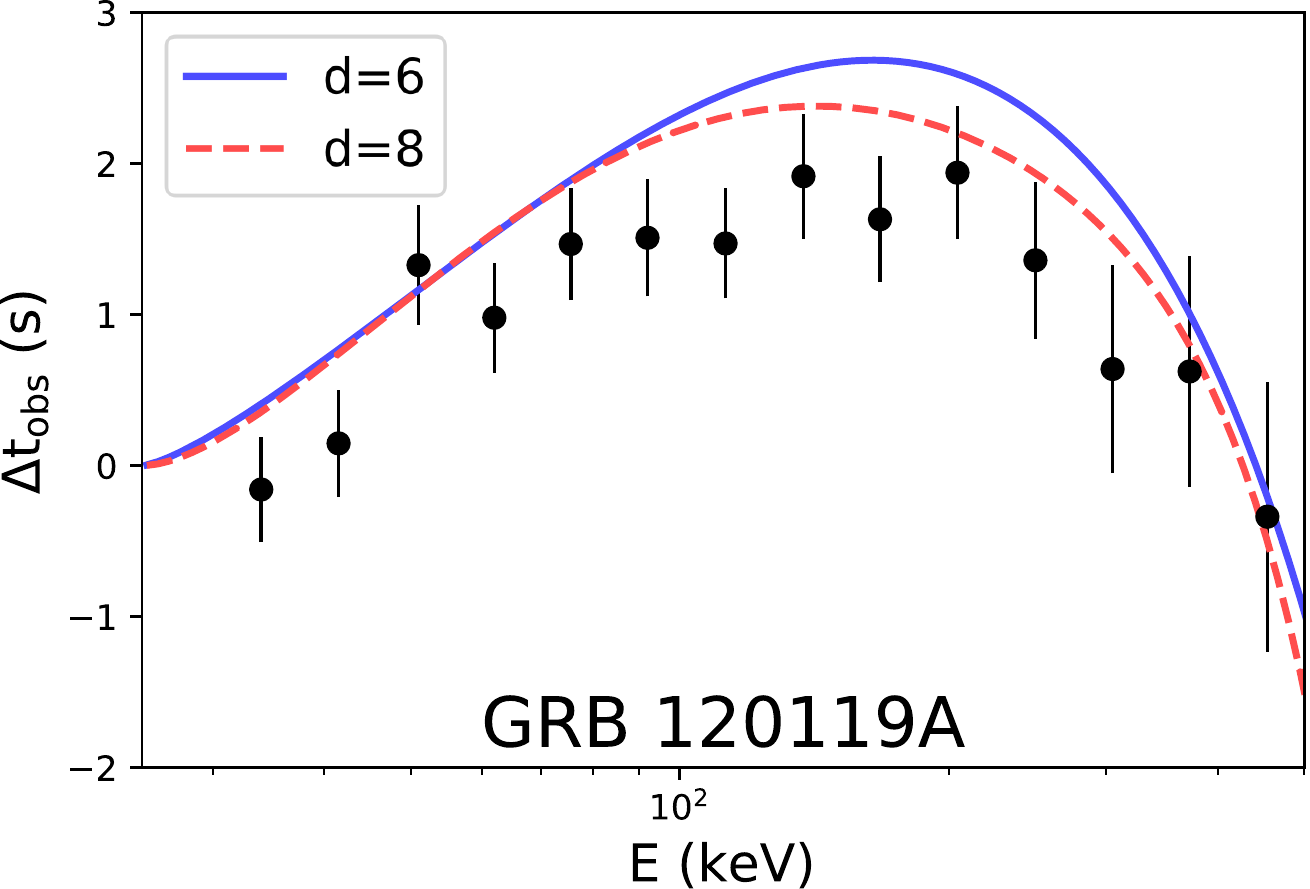}}
\subfigure{\includegraphics[width=4.2cm]{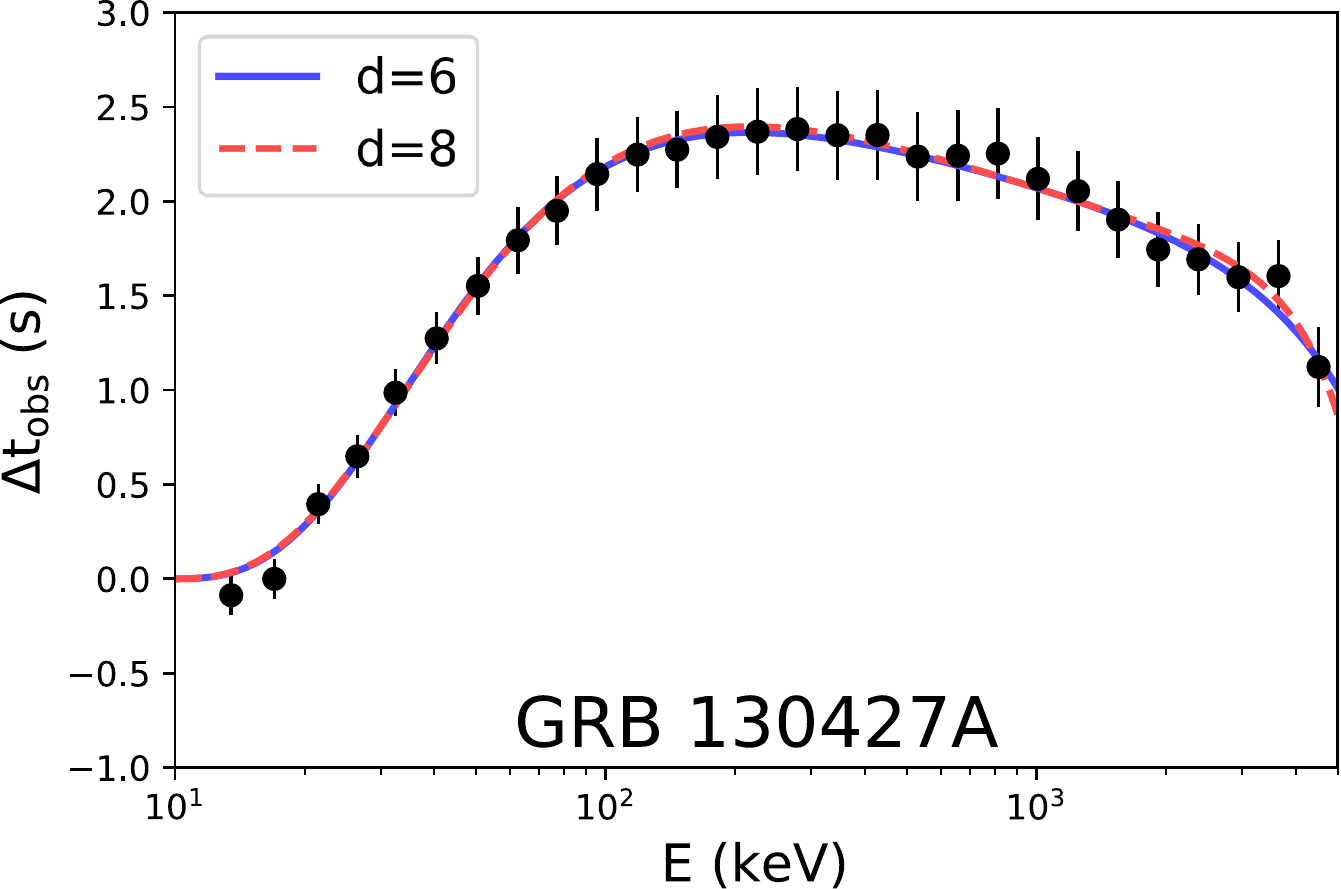}}
\subfigure{\includegraphics[width=4.2cm]{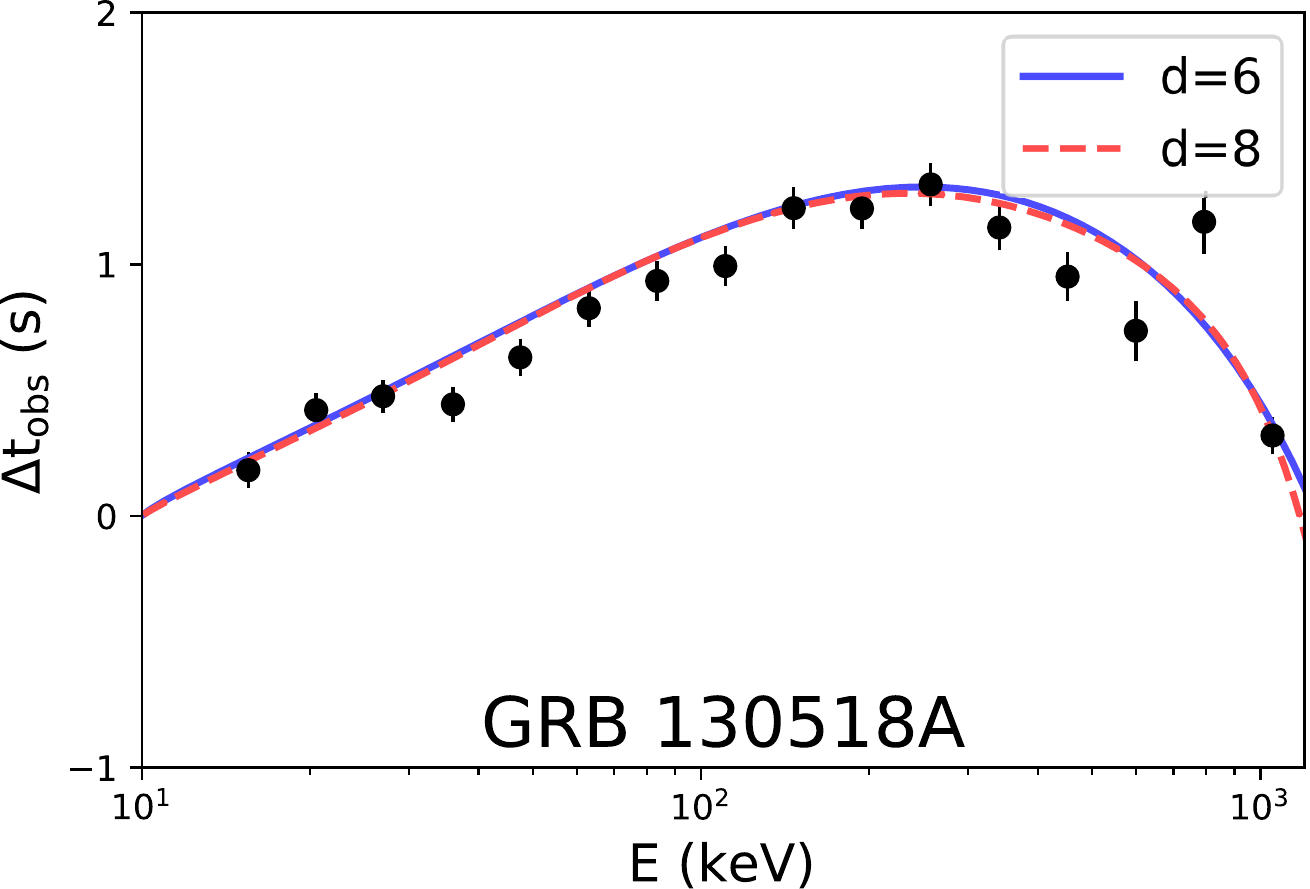}}
\subfigure{\includegraphics[width=4.2cm]{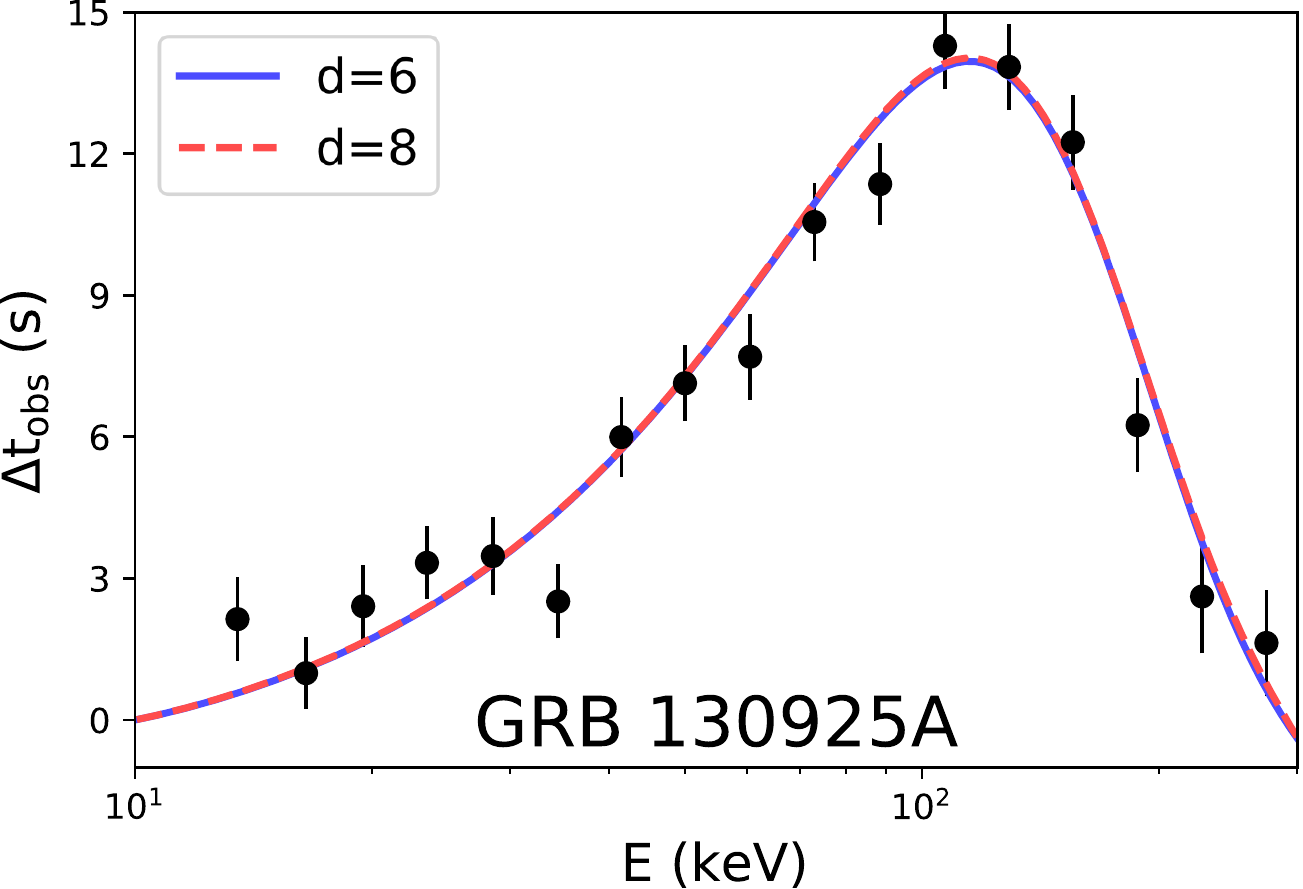}}
\subfigure{\includegraphics[width=4.2cm]{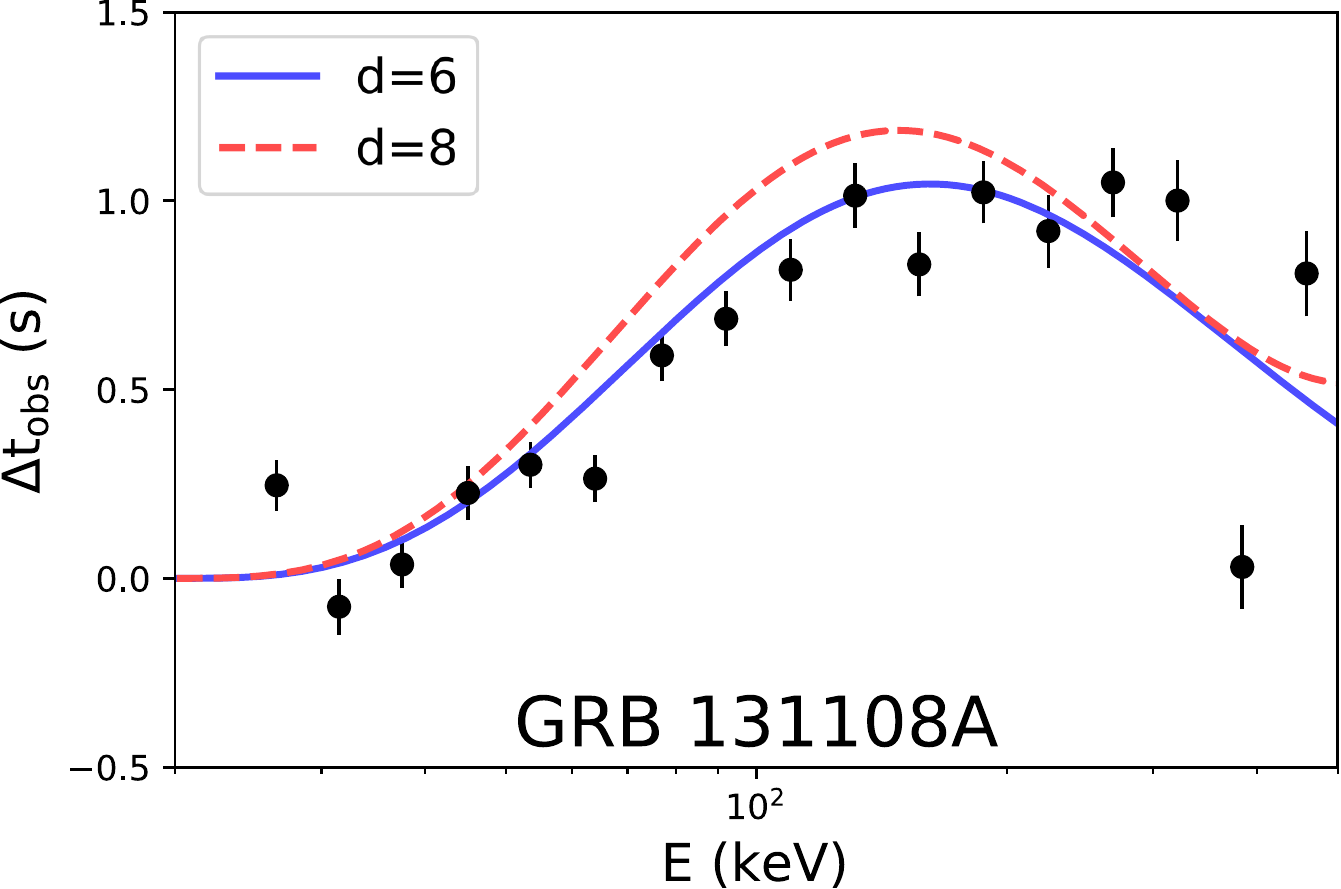}}
\subfigure{\includegraphics[width=4.2cm]{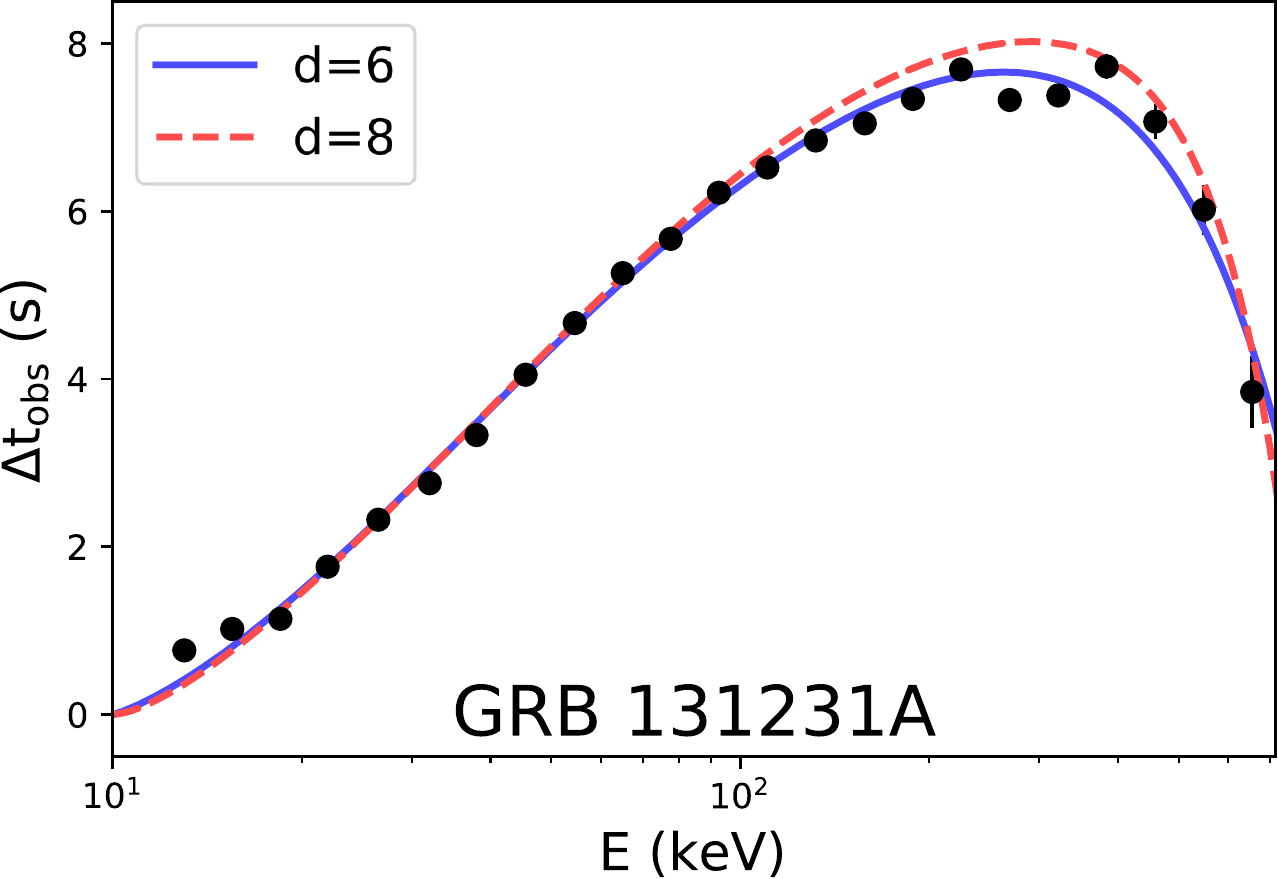}}
\subfigure{\includegraphics[width=4.2cm]{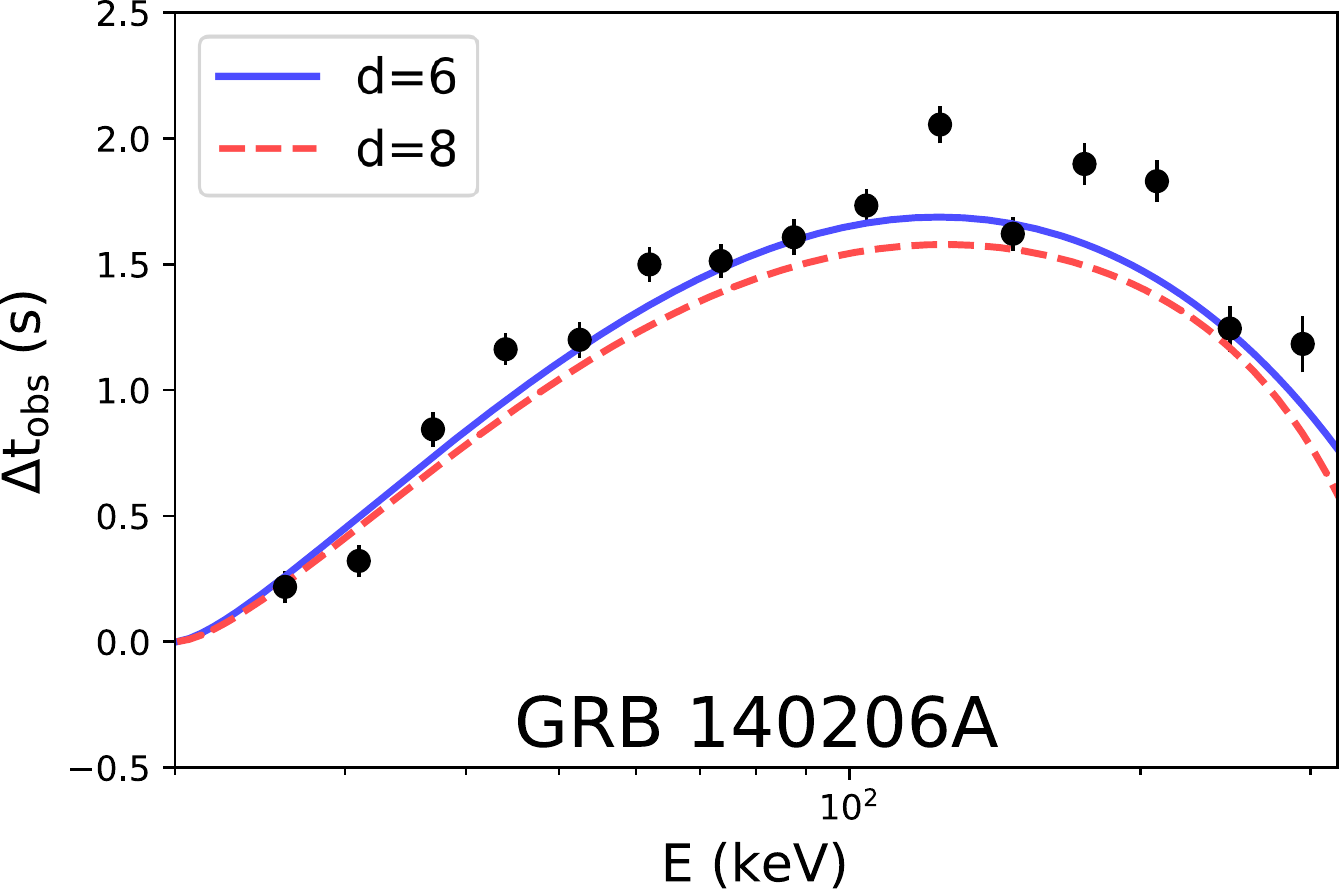}}
\subfigure{\includegraphics[width=4.2cm]{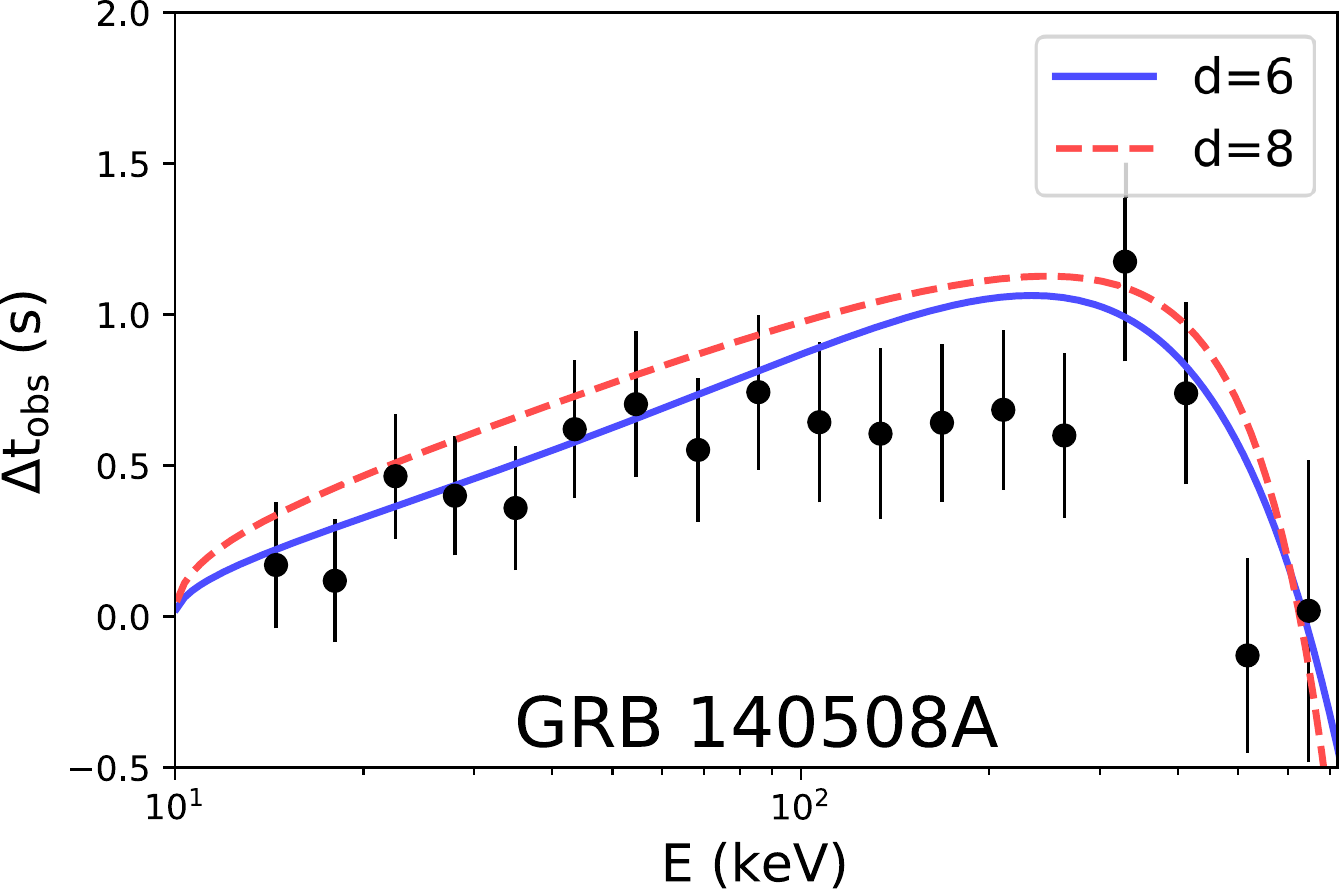}}
\subfigure{\includegraphics[width=4.2cm]{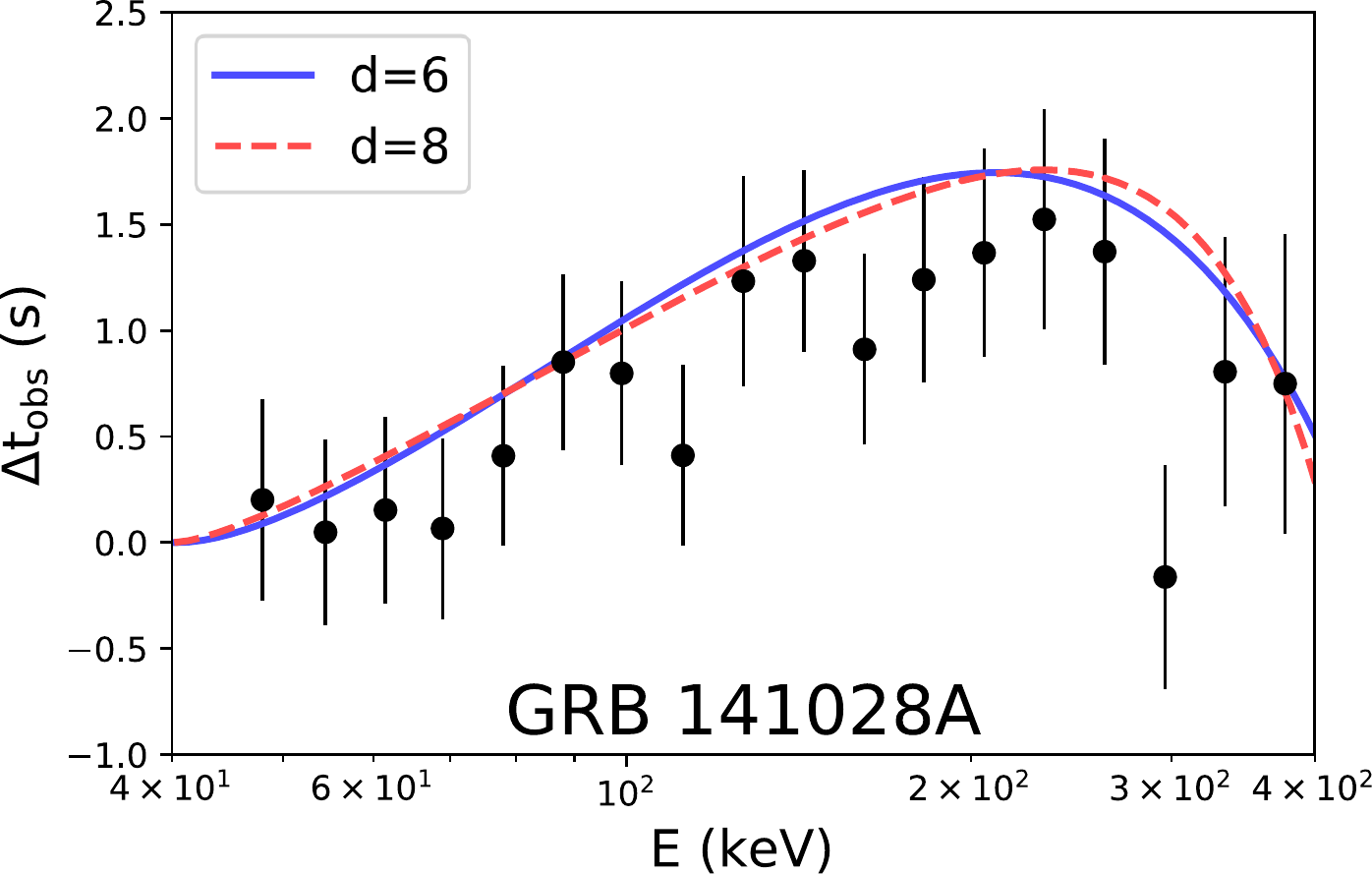}}
\subfigure{\includegraphics[width=4.2cm]{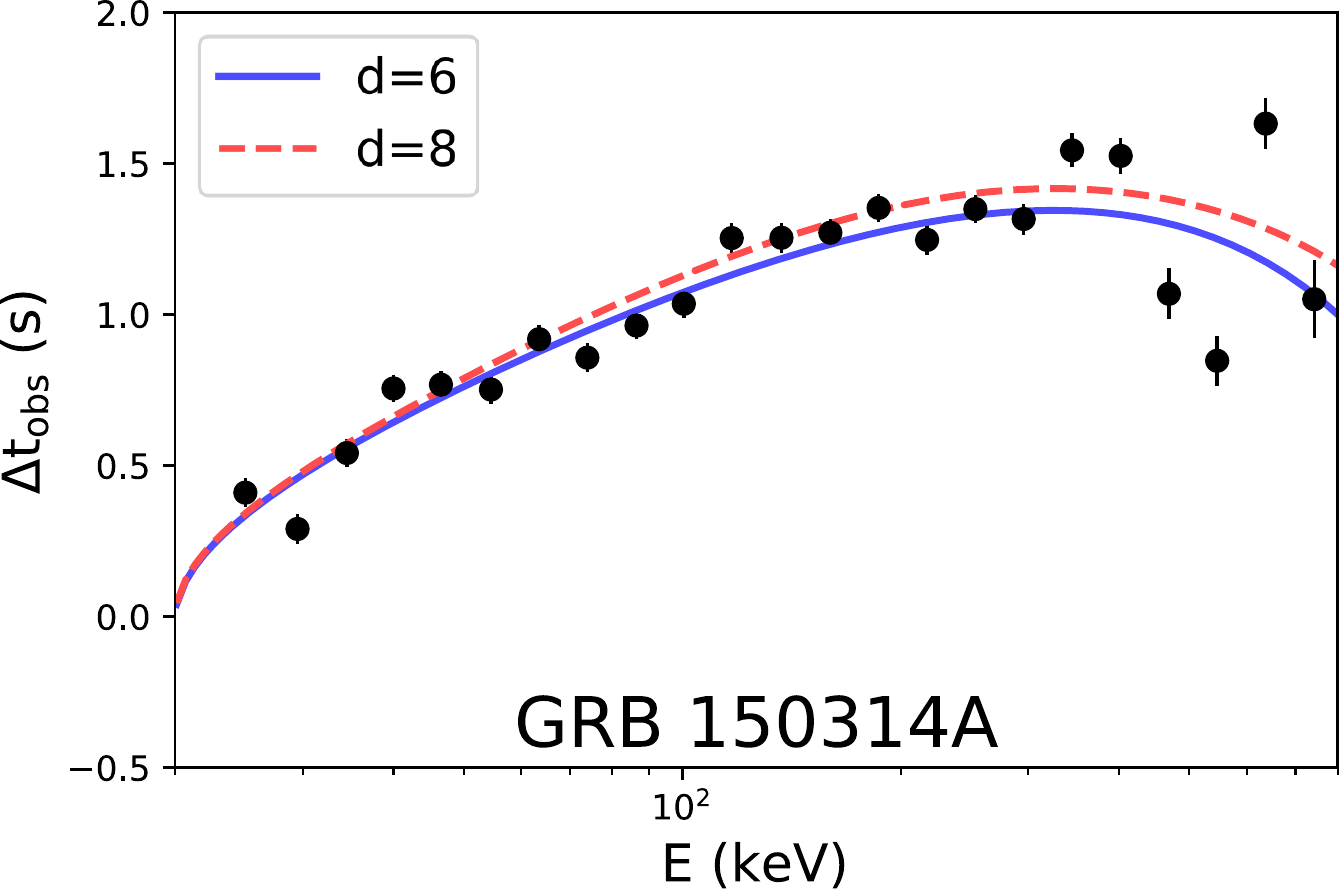}}
\subfigure{\includegraphics[width=4.2cm]{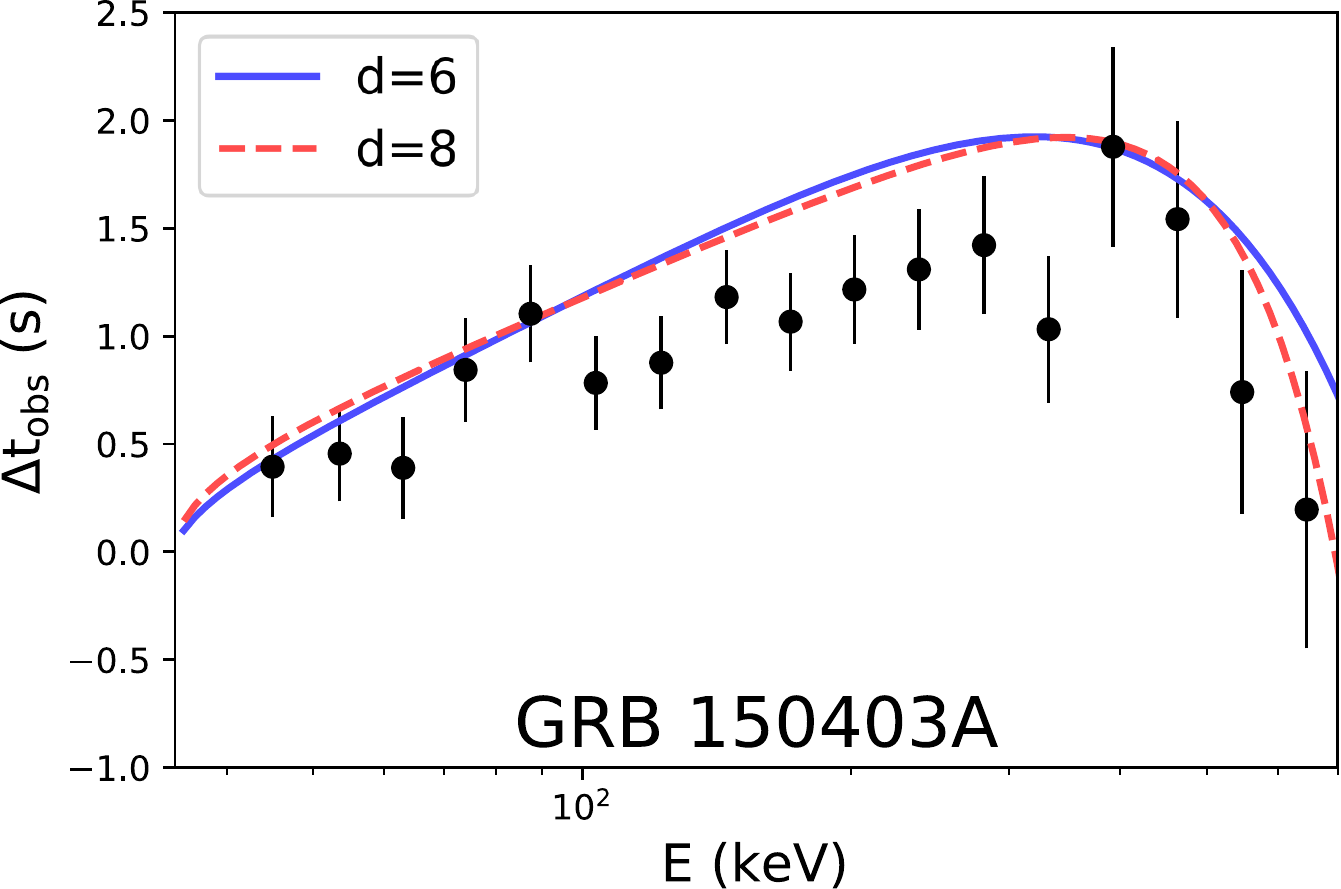}}
\caption{\textit{Cont}.}
\end{figure}

\begin{figure}[H]\ContinuedFloat
\subfigure{\includegraphics[width=4.6cm]{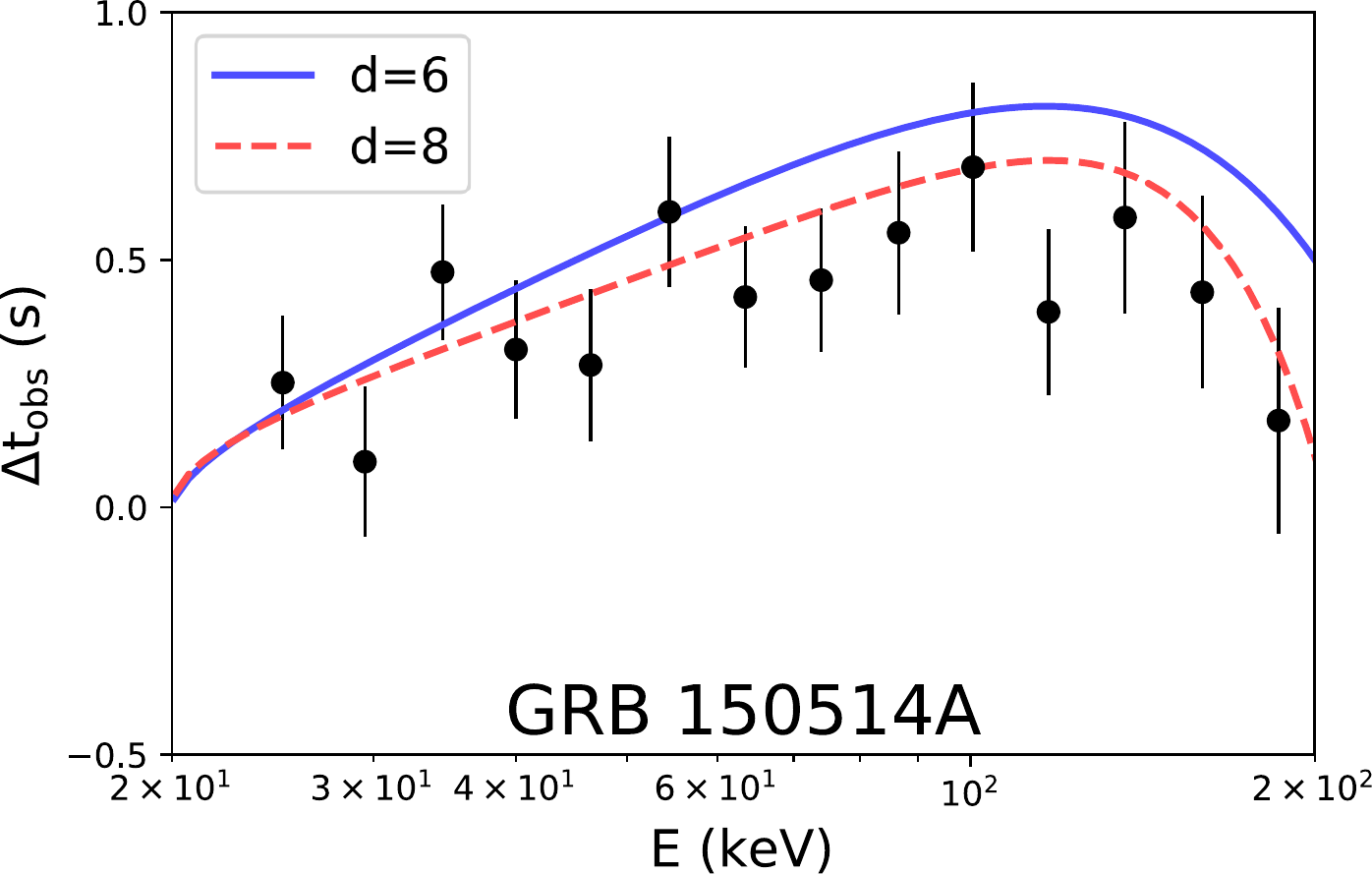}}
\subfigure{\includegraphics[width=4.15cm]{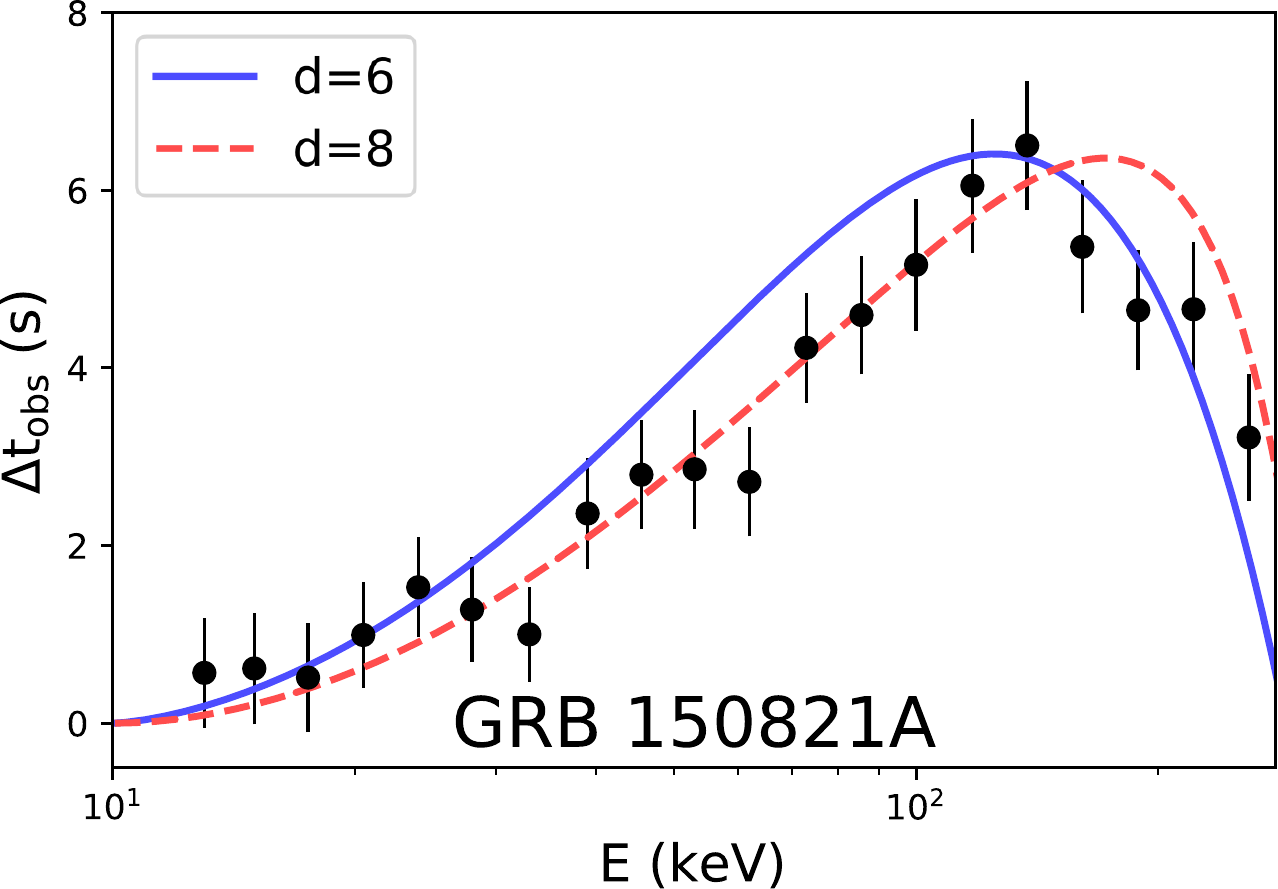}}
\subfigure{\includegraphics[width=4.2cm]{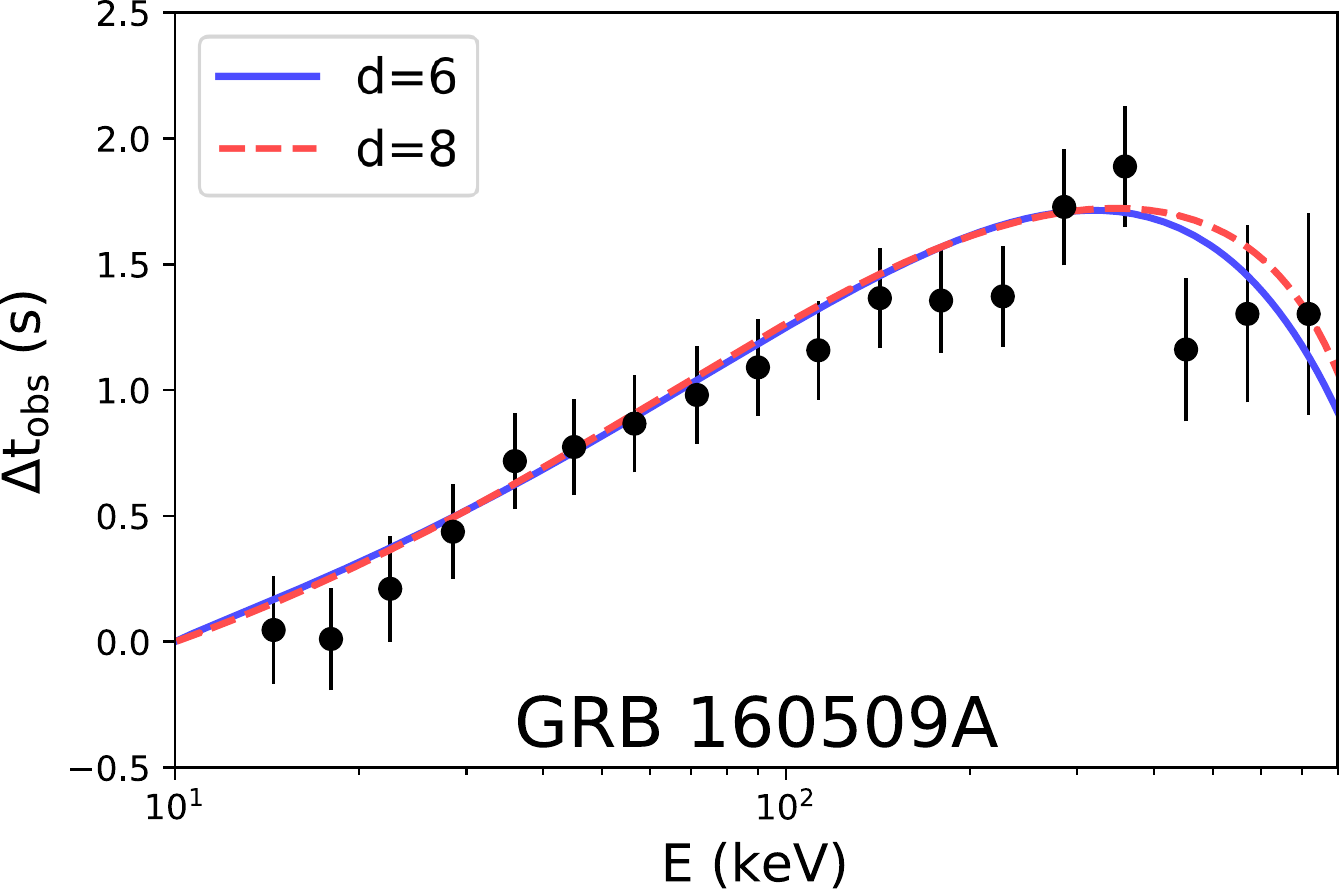}}
\subfigure{\includegraphics[width=4.2cm]{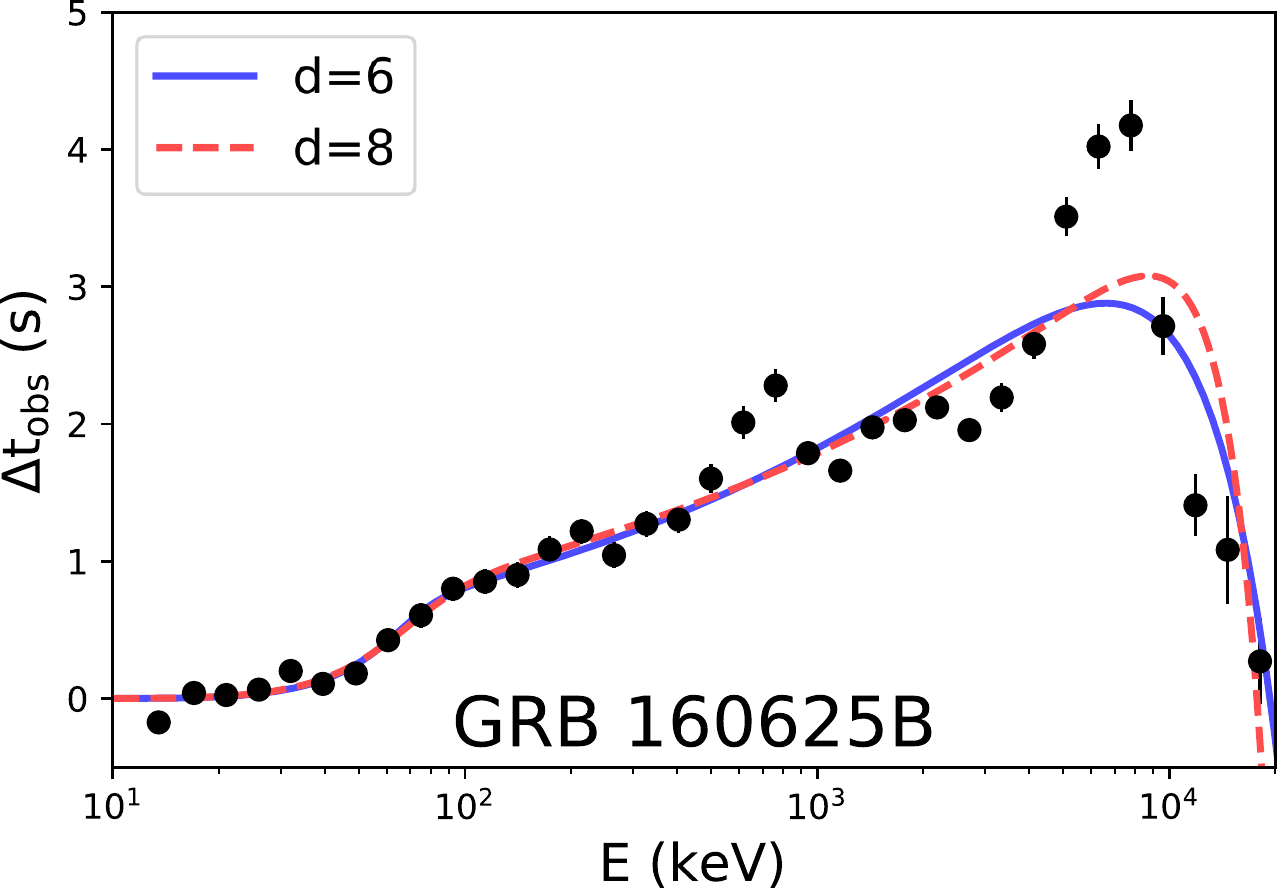}}
\subfigure{\includegraphics[width=4.2cm]{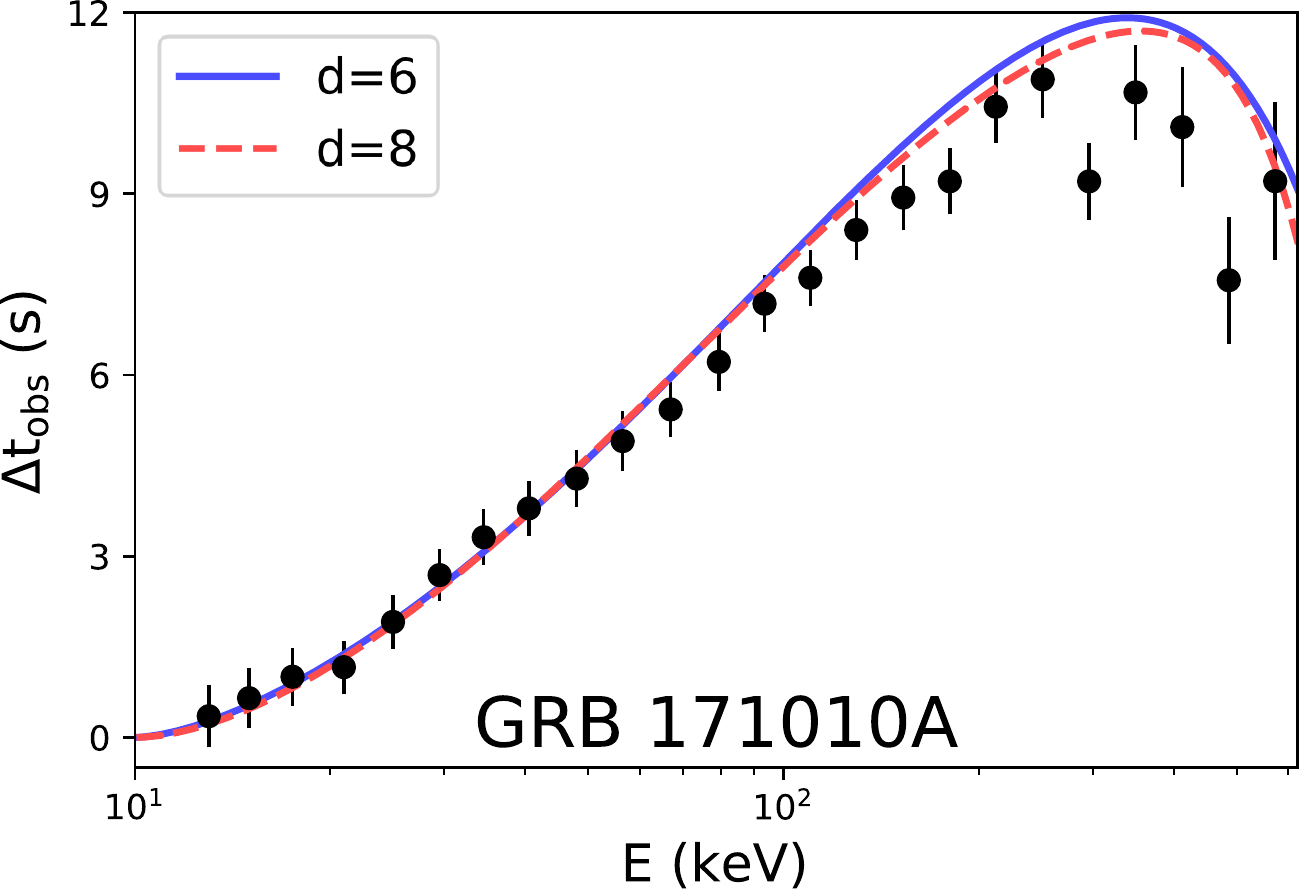}}
\subfigure{\includegraphics[width=4.2cm]{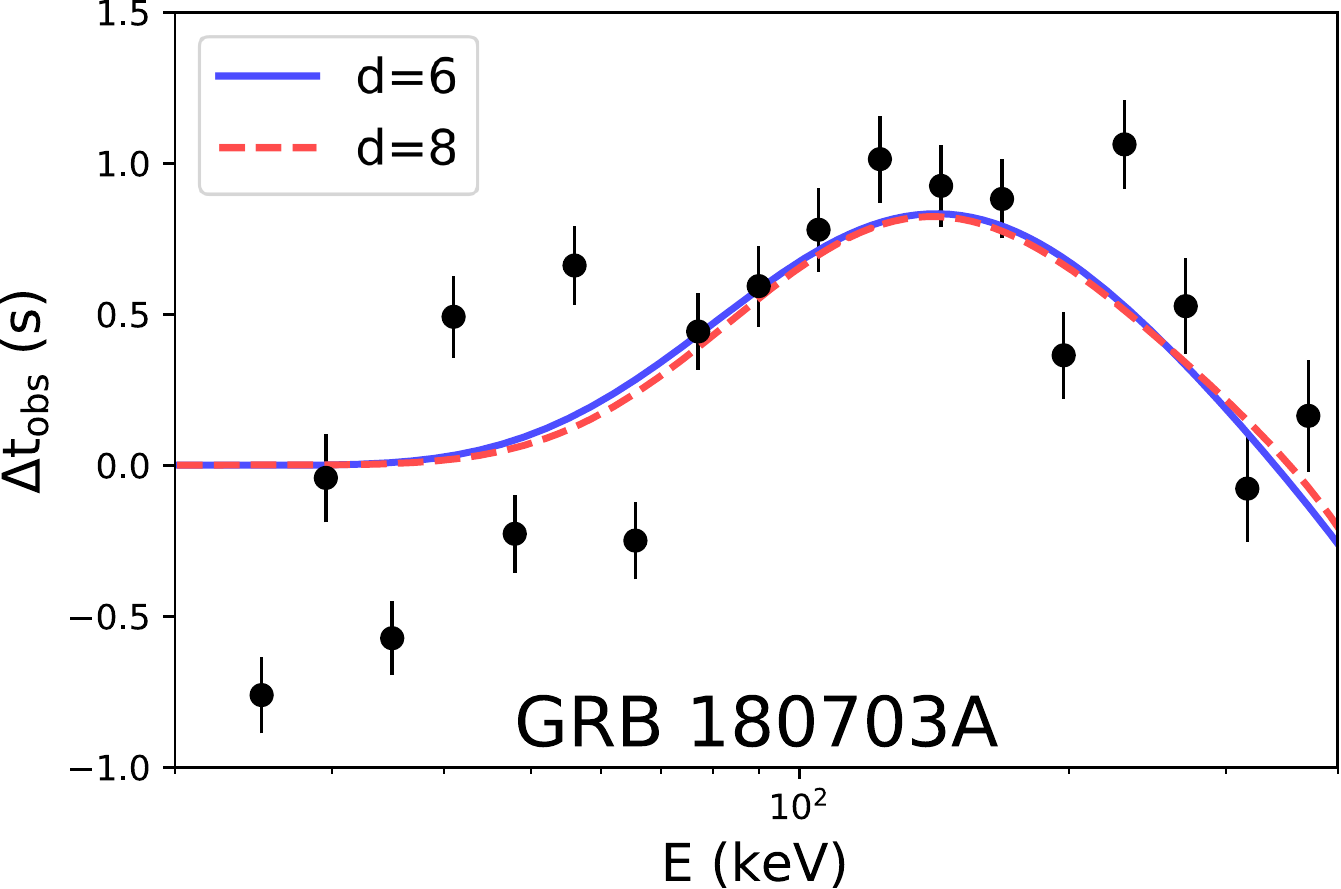}}
\subfigure{\includegraphics[width=4.2cm]{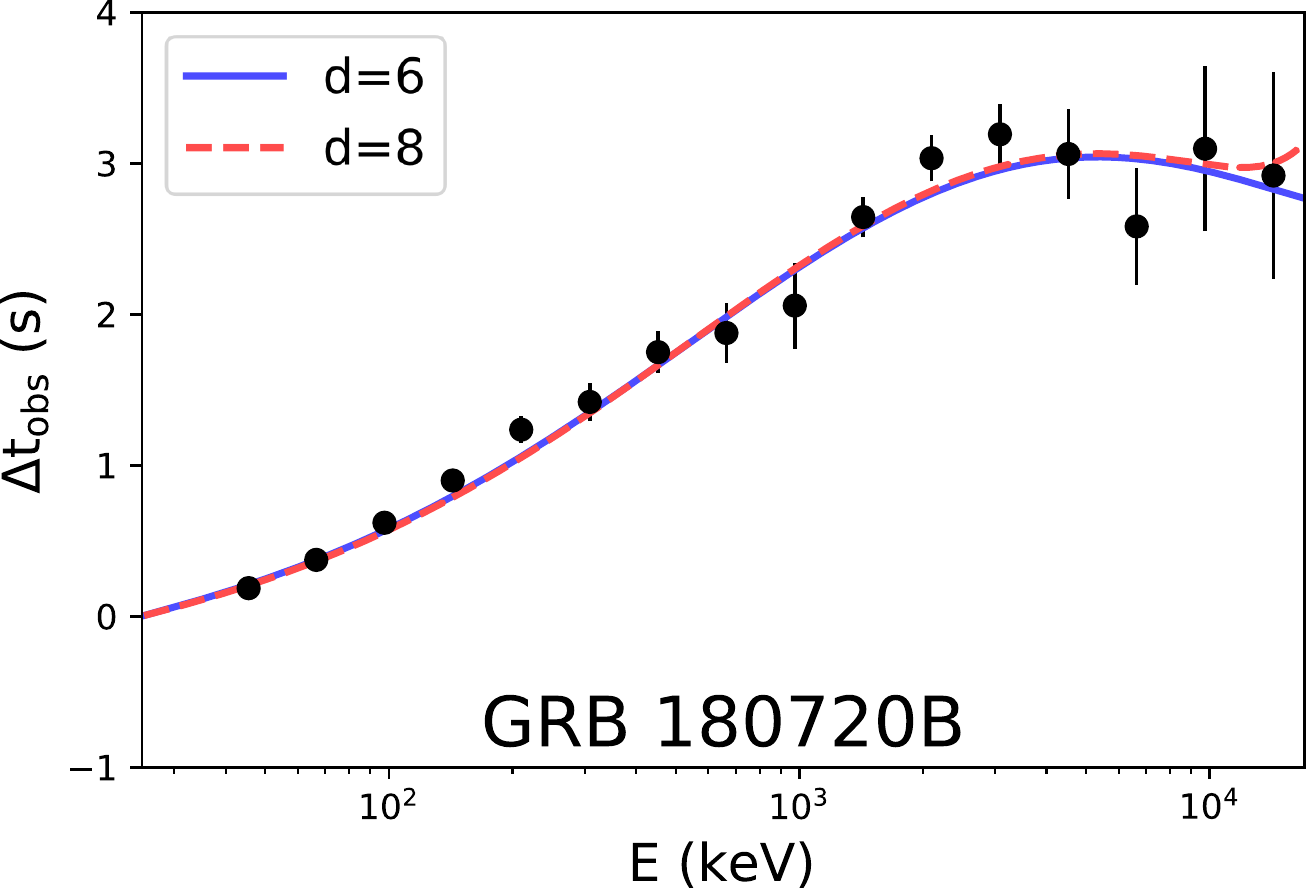}}
\subfigure{\includegraphics[width=4.2cm]{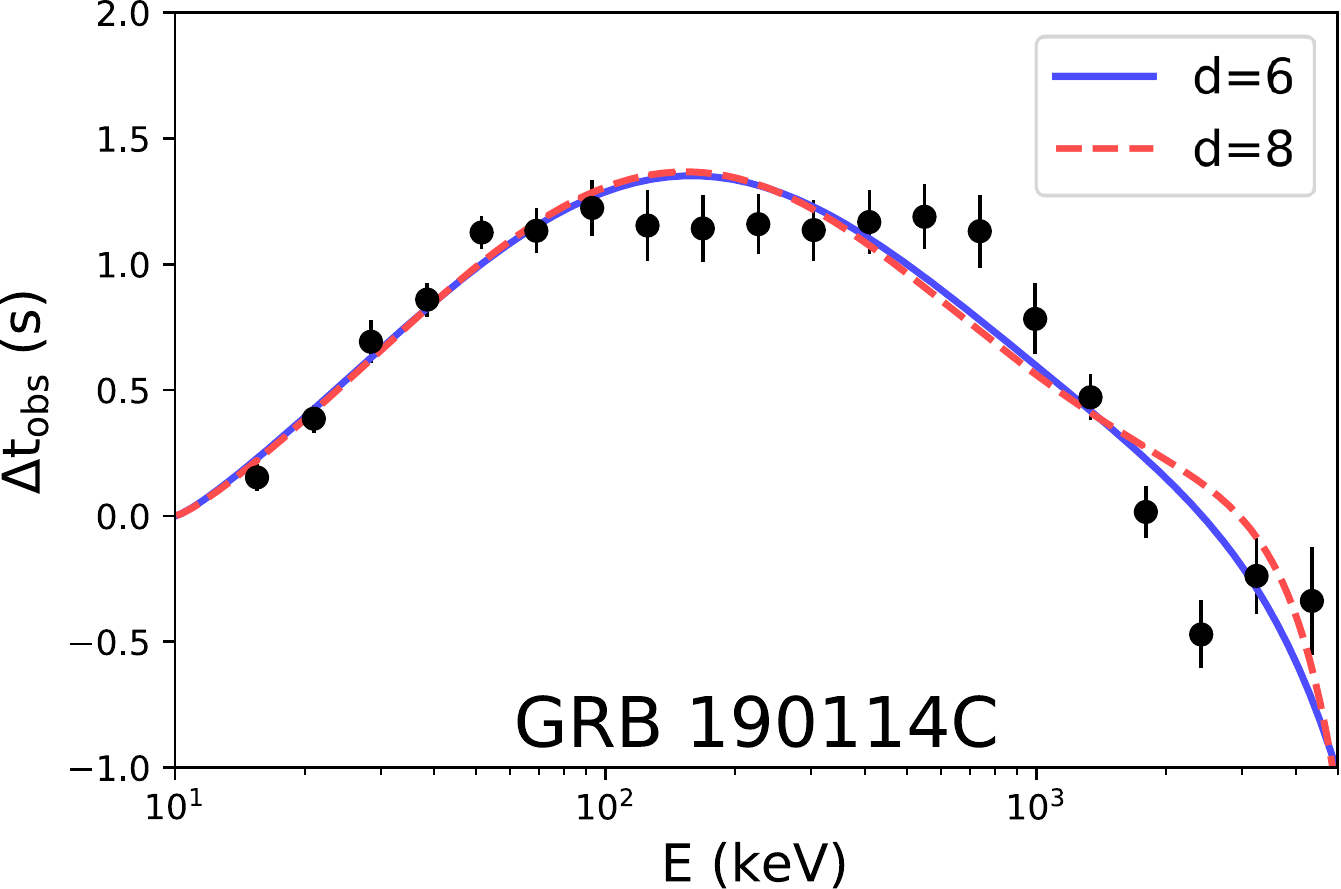}}
\subfigure{\includegraphics[width=4.35cm]{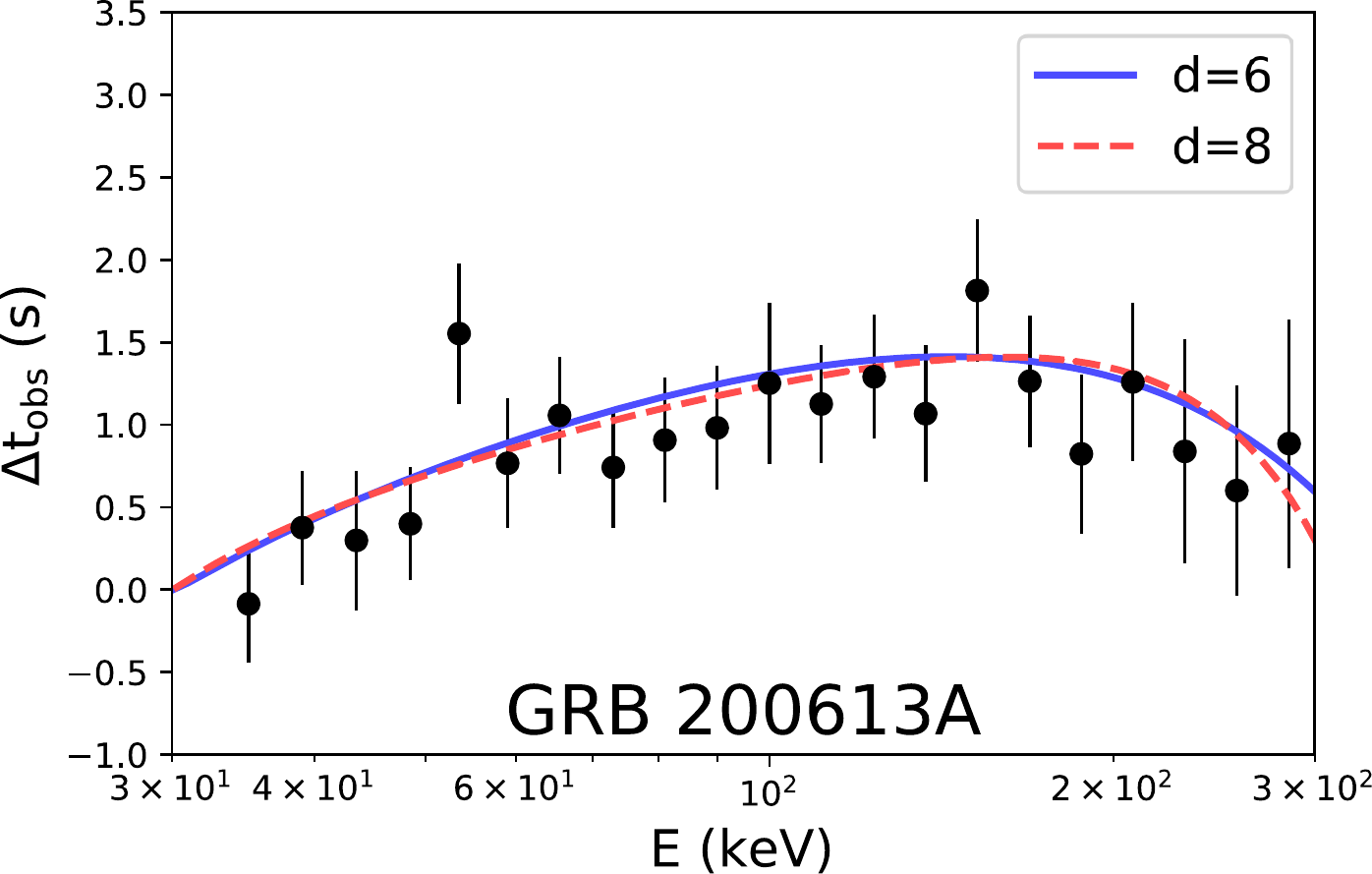}}
\subfigure{\includegraphics[width=4.2cm]{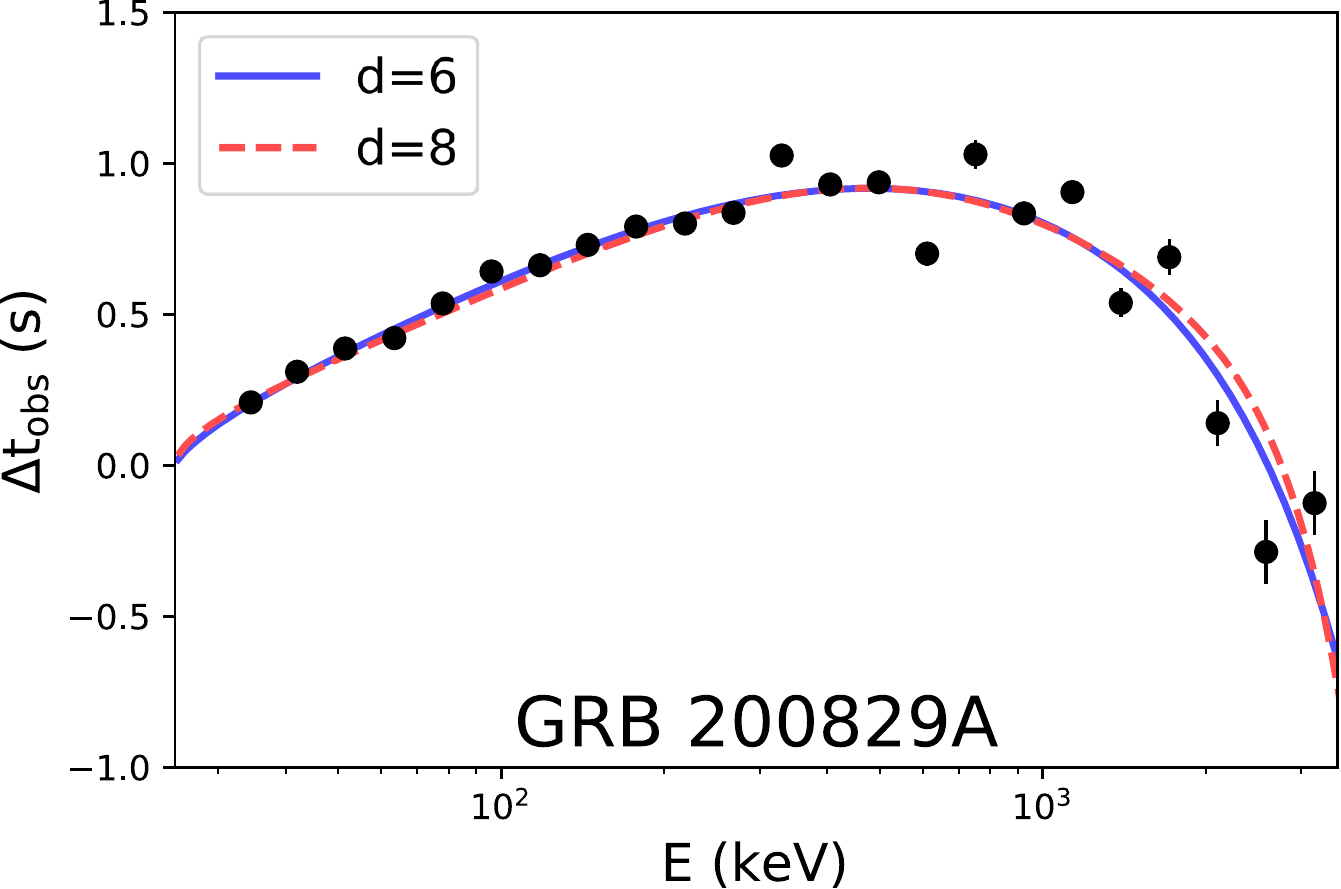}}
\subfigure{\includegraphics[width=4.2cm]{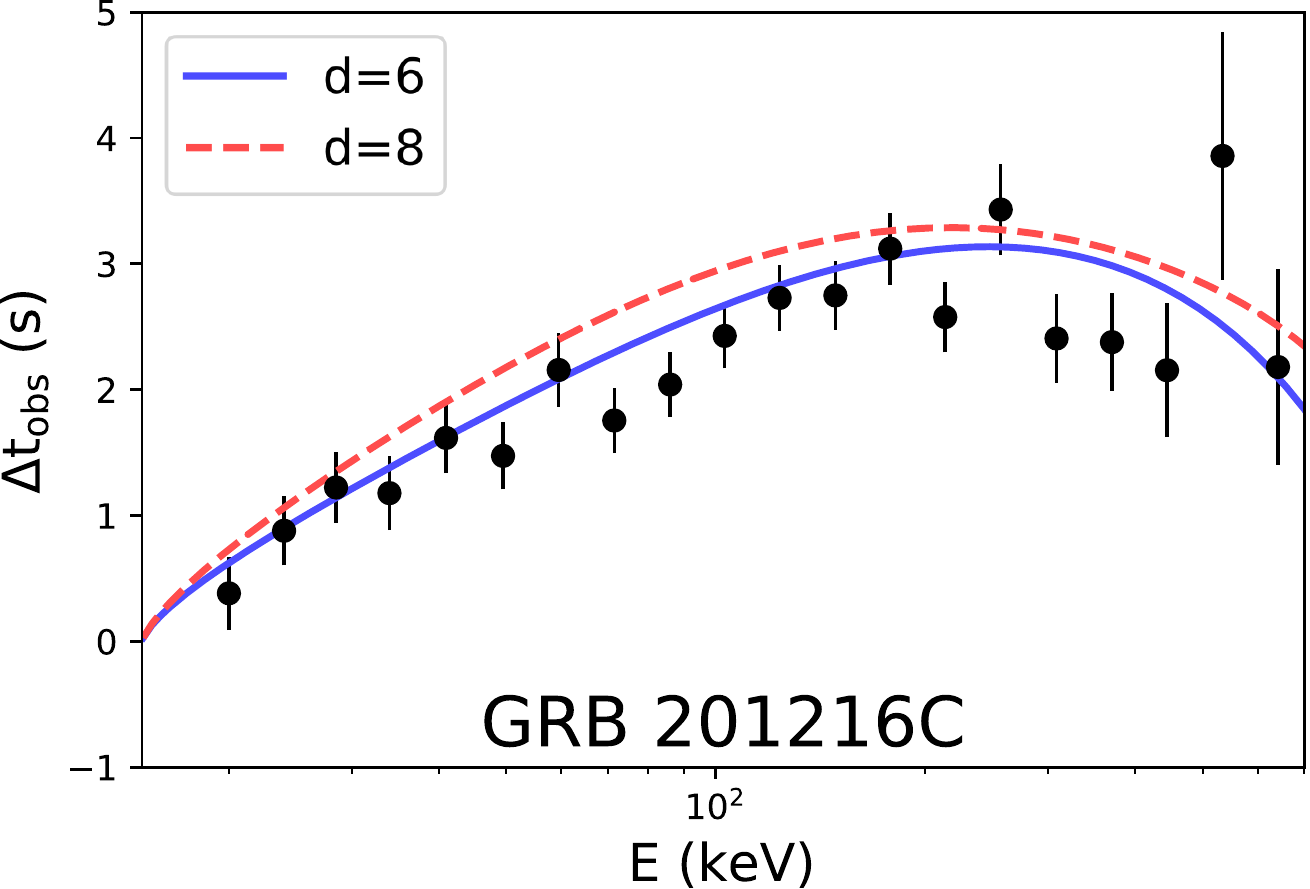}}
\subfigure{\includegraphics[width=4.2cm]{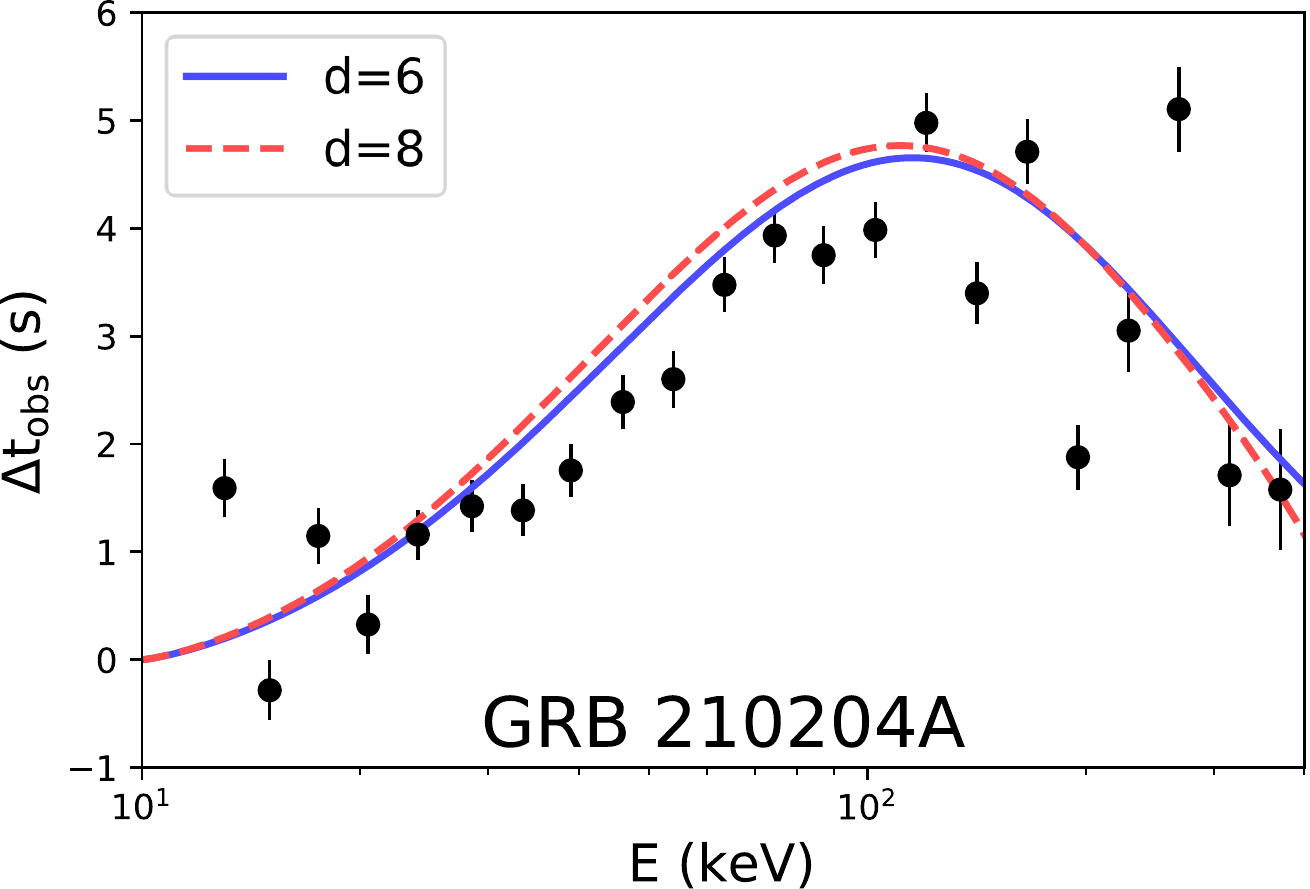}}
\subfigure{\includegraphics[width=4.2cm]{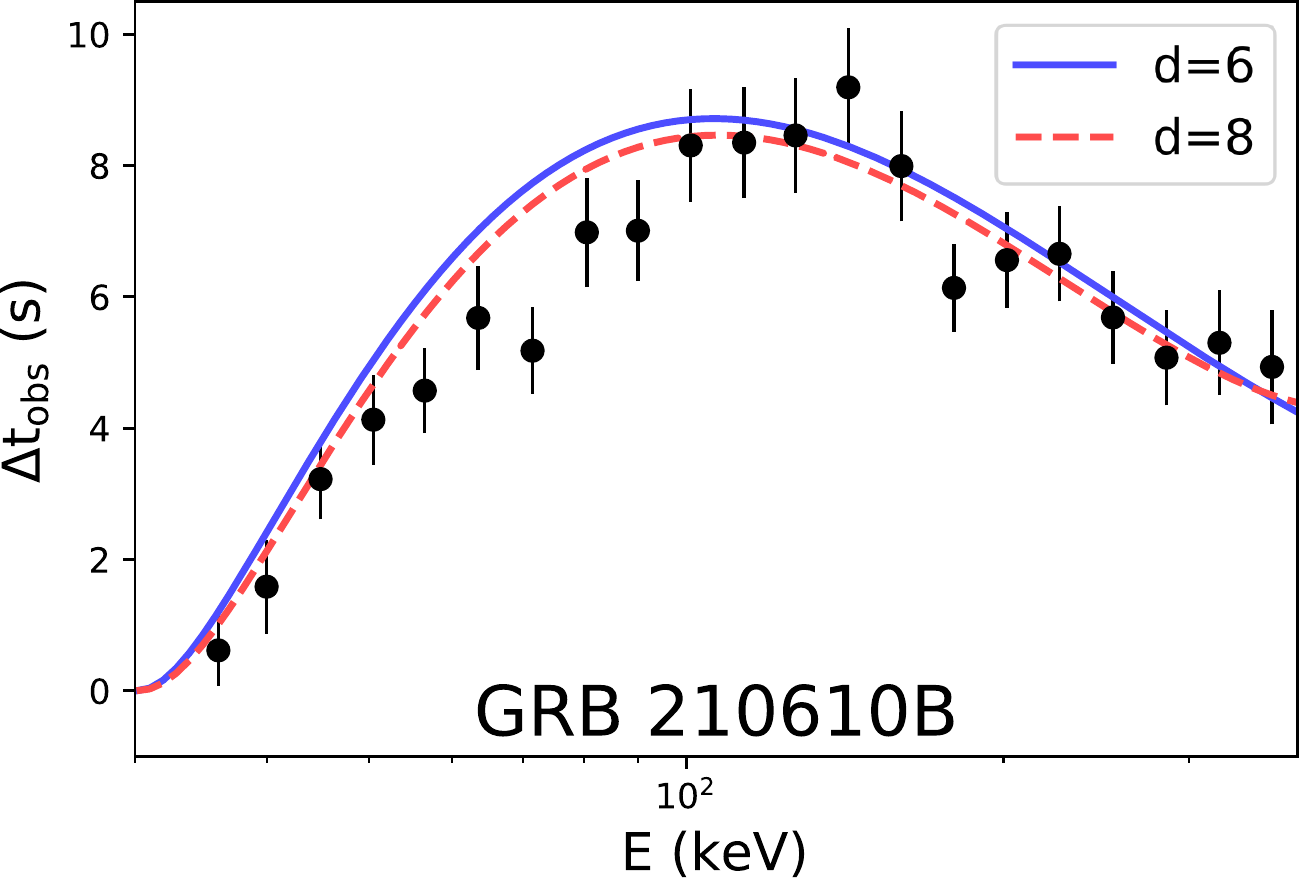}}
\subfigure{\includegraphics[width=4.2cm]{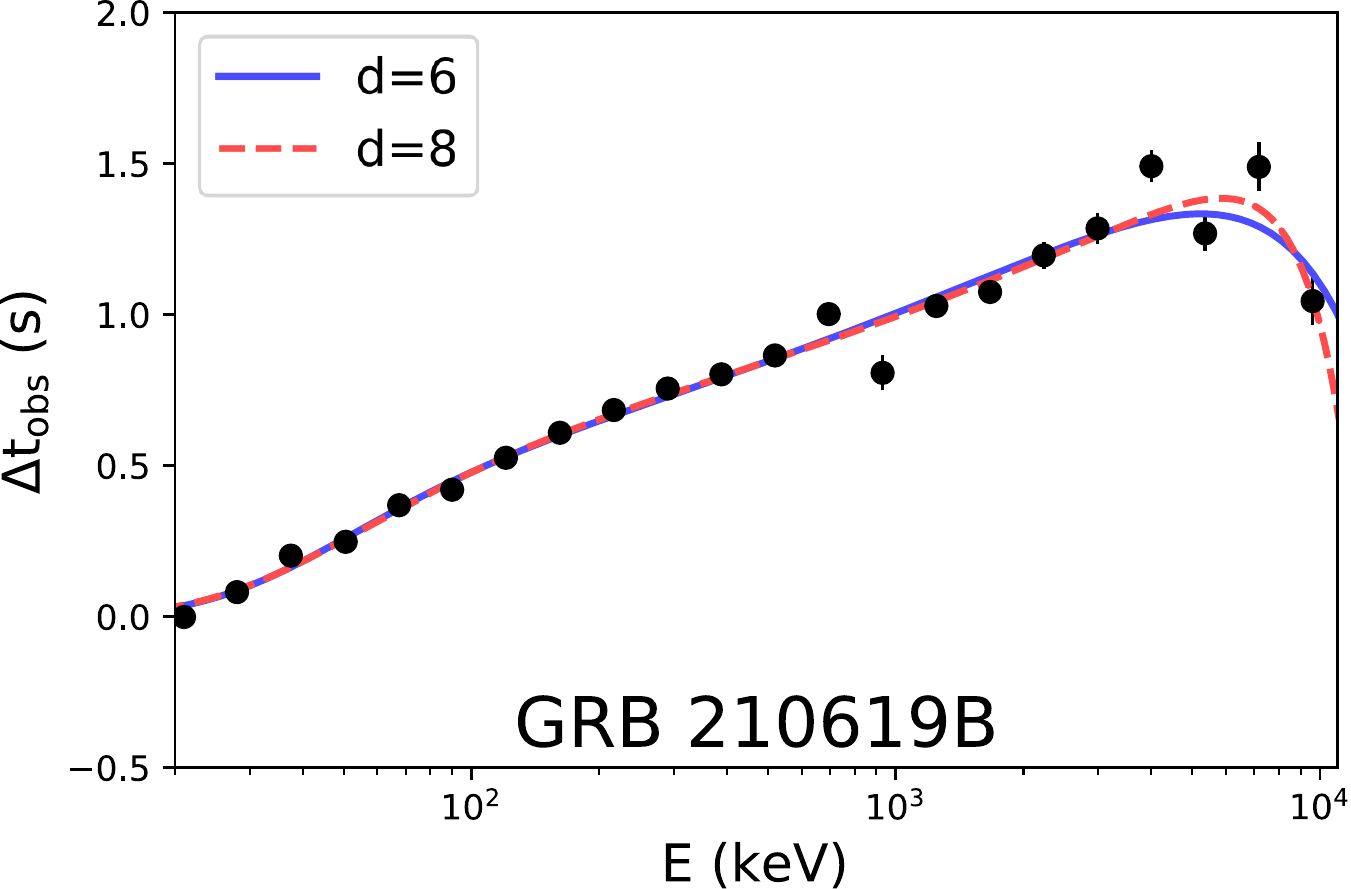}}
\caption{Energy dependence of the observed spectral lag $\Delta t_{\rm obs}$ of each GRB in our sample, 
and the best-fit theoretical curves. Blue solid lines: model with $d=6$ coefficients. Red dashed lines: 
model with $d=8$ coefficients.\label{fig1}}
\end{figure}

In view of the LIV-induced time delay, $\Delta t_{\rm LIV}$ is likely to be accompanied by an intrinsic 
time delay $\Delta t_{\rm int}$ caused by the emission mechanism of GRBs 
\citep{2006APh....25..402E,2009CQGra..26l5007B,2017ApJ...834L..13W,2017ApJ...842..115W}; 
the observed spectral lag should consist of two terms 
\begin{equation}
\Delta t_{\rm obs} = \Delta t_{\rm int} + \Delta t_{\rm LIV}\;.
\label{eq:tobs}
\end{equation}
In principle, 
it is hard to model the intrinsic lag behavior because of the unknown GRB emission mechanism. 
As previous works have excluded some nonbirefringent Lorentz-violating effects with a high level of confidence 
(e.g., \citep{2013PhRvD..87l2001V}), it is reasonable to assume that the observed spectral lag is dominated by 
the intrinsic time lag, and the LIV-induced time lag is negligible. Statistically, the time lags of these 32 GRBs 
first increase and then decrease with the energies in the form of an approximate broken power-law function, 
i.e., the broken power-law model is in fact an accurate representation of the energy dependence of the intrinsic 
time lag. Hence, as Liu et al. \cite{2022arXiv220209999L} conducted in their treatment, we adopt a smoothly broken 
power-law (SBPL) function to model the intrinsic energy-dependent time lag,
\begin{equation}
\Delta t_{\rm int} = \tau \left(\frac{E - E_{\rm l}}{E_{b}}\right)^{\alpha_{1}} \left\{ \frac{1}{2}\left[1 + \left(\frac{E - E_{\rm l}}{E_{b}}\right)^{1/\beta}\right]\right\}^{(\alpha_{2} - \alpha_{1})\beta}\;,
\label{eq:tint}
\end{equation}
where $\tau$ is the normalization amplitude, $E_{\rm l}$ is the median value of the lowest energy band 
of each GRB, $\alpha_1$ and $\alpha_2$ are the power-law indices before and after the break energy
$E_{b}$, and $\beta$ measures the smoothness of the transition. In order to ensure that 
$\Delta t_{\rm int}$ is consistent with the observed lag behavior displaying a transition from positive to
negative lags, we require the $\Delta t_{\rm LIV}$ term in Equation~(\ref{eq:tobs}) not to dominate over
$\Delta t_{\rm int}$. That is, we require $\alpha_1\geq\alpha_2$.

We fit the lag-energy measurements of each GRB using the theoretical model as shown in
Equations~(\ref{eq:tLIV})--(\ref{eq:tint}). The free parameters ($\tau$, $\alpha_{1}$, $\alpha_{2}$, $E_{b}$, and
$\beta$) of the SBPL function and the combination $\sum_{jm} {}_{0}Y_{jm}(\hat{\textbf{\emph{n}}}) c_{(I)jm}^{(d)}$ 
of SME coefficients are simultaneously estimated by using the Markov chain Monte Carlo (MCMC) technique.
The Python MCMC module EMCEE~\citep{2013PASP..125..306F} is adopted to explore the posterior probability
distributions of these parameters. The log-likelihood sampled by EMCEE is given by
\begin{equation}
\ln L(\theta) = -\frac{1}{2}\sum_{i}\frac{\left[\Delta t_{{\rm obs},i} - \Delta t_{\rm model}(\theta)\right]^{2}}{\sigma^{2}_{\Delta t_{{\rm obs},i}}}\;,
\end{equation}
where $\sigma_{\Delta t_{{\rm obs},i}}$ is the uncertainty of the $i$th lag measurement 
$\Delta t_{{\rm obs},i}$, $\Delta t_{\rm model}(\theta)$ is obtained from Equation~(\ref{eq:tobs}),
and $\theta$ denotes the free parameters. In our analysis, we choose wide flat priors for $\tau\in[0.0,\;4.0]$ s,
$\alpha_{1}\in[-3.0,\;10.0]$, $\alpha_{2}\in[-10.0,\;3.0]$, $\beta\in[0.0,\;3.0]$, and $E_{b}\in[0.0,\;5000.0]$ keV. 
The prior of the combination $\sum_{jm} {}_{0}Y_{jm}(\hat{\textbf{\emph{n}}}) c_{(I)jm}^{(d)}$ of SME coefficients 
is set as a uniform distribution in the range of [$-10^{-10}$, $10^{-10}$] ${\rm GeV}^{-2}$ for $d=6$, 
or [$-10^{-3}$, $10^{-3}$] ${\rm GeV}^{-4}$ for $d=8$. Note that the spectral-lag transition from positive 
to negative is for the power-law indices ($\alpha_1$ and $\alpha_2$) when fitting the observed lag-energy data.

\begin{figure}[H]
\subfigure{\includegraphics[width=11.5cm]{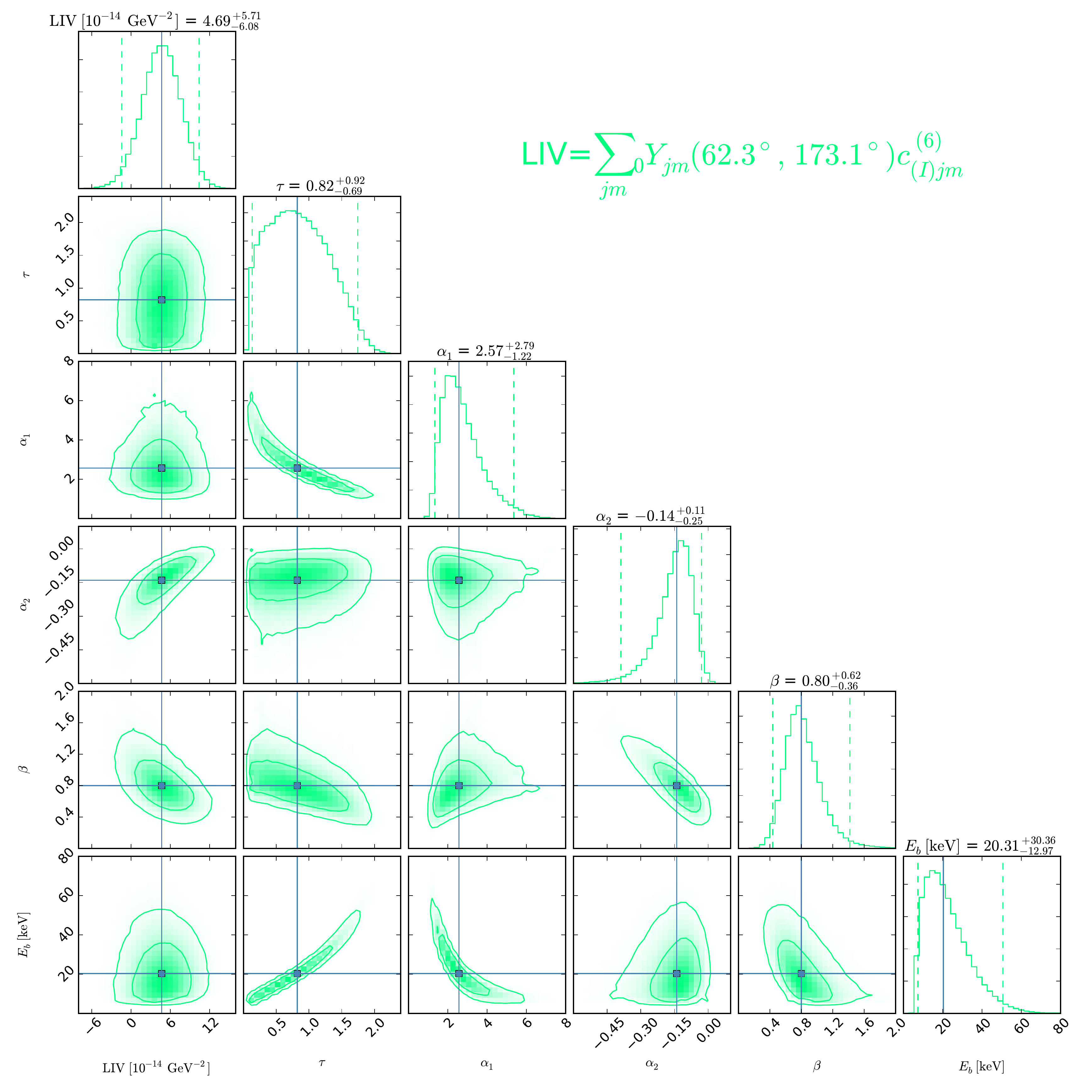}}
\subfigure{\includegraphics[width=11.5cm]{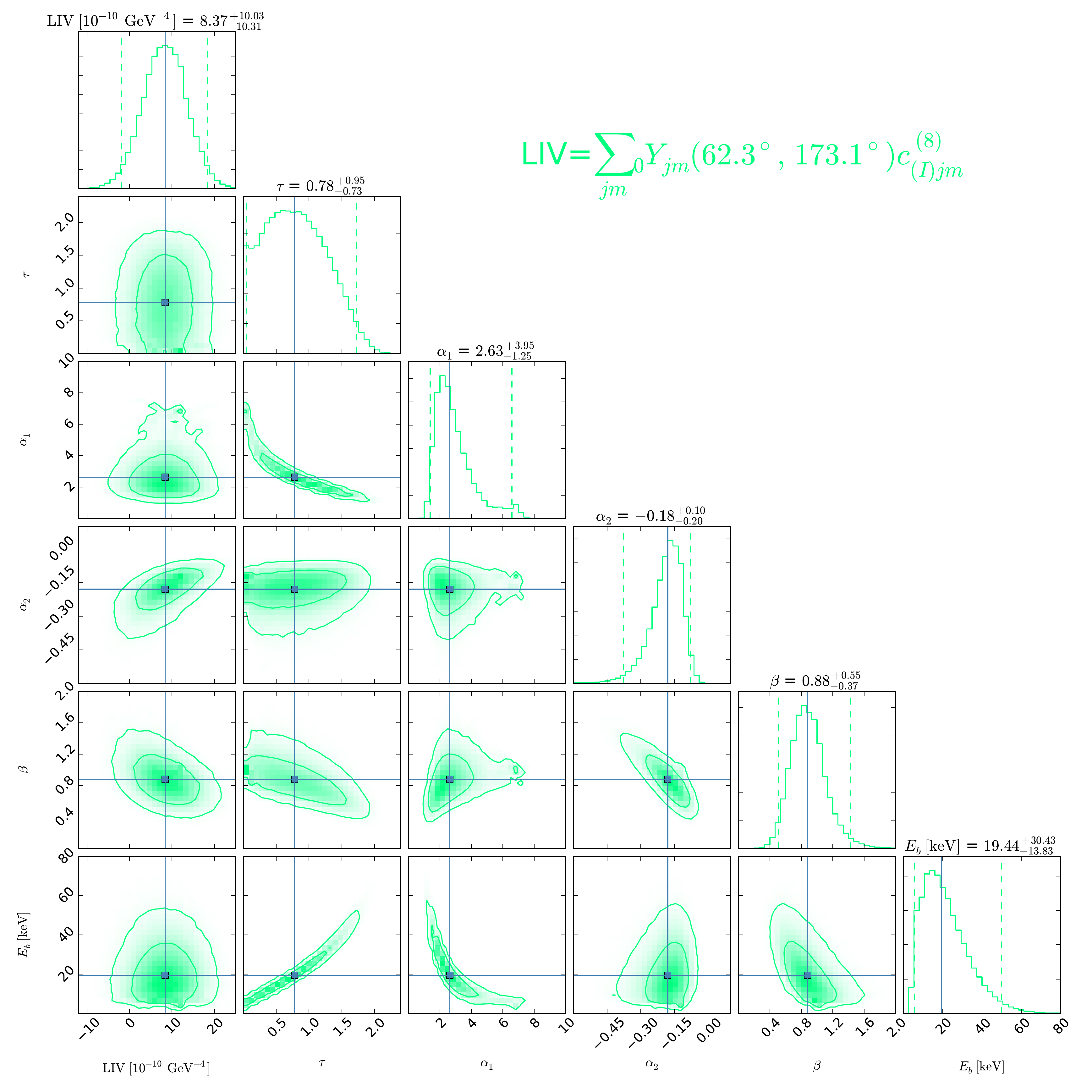}}
\caption{Upper panel: 1D and 2D posterior probability distributions with the 1-2$\sigma$ contours for the}
\label{fig2}
\end{figure}

The resulting constraints from the observed lag-energy data of each GRB are presented in Table~\ref{tab2}. 
For each value $d=6,8$ in turn, the best-fit results and 2$\sigma$ uncertainties are provided for 
the parameters $\tau$, $\alpha_{1}$, $\alpha_{2}$, and $E_{b}$ and for the direction-dependent combination 
$\sum_{jm} {}_{0}Y_{jm}(\hat{\textbf{\emph{n}}}) c_{(I)jm}^{(d)}$ of Lorentz-violating coefficients.
Table~\ref{tab2} also provides estimated constraints at the 95\% confidence level for the corresponding 
coefficients in the restrictive isotropic limit $j=m=0$. To illustrate the goodness of fits, 
the theoretical curves obtained from the best-fit values are plotted in Figure~\ref{fig1} for models
with $d=6$ and $d=8$ coefficients. Moreover, Figure~\ref{fig2} shows an example of the posterior probability
distributions of the free parameters for GRB 130427A. 
The probability distributions for models with
$d=6$ and $d=8$ coefficients are presented in the upper and lower panels of Figure~\ref{fig2}, respectively. 
The corresponding distributions for other GRBs are qualitatively similar. 
\begin{table}[H]
\caption{95\% C.L. constraints on the SME coefficients and the parameters of the SBPL function.\label{tab2}}
\begin{adjustwidth}{-\extralength}{0cm}
\tablesize{\scriptsize}
\begin{tabularx}{\fulllength}{lCm{0.5cm}m{2.5cm}m{2.5cm}CCCC}
\toprule
\multirow{3}{*}{\textbf{Name}} & \multirow{3}{*}{\boldmath{$(\theta,\;\phi)$}} & \multirow{3}{*}{\boldmath{$d$}} & \boldmath{$\sum_{jm} {}_{0}Y_{jm}(\theta,\;\phi) c_{(I)jm}^{(d)}$} & \boldmath{$c_{(I)00}^{(d)}$} & \multirow{3}{*}{\boldmath{$\tau$}}  &  \multirow{3}{*}{\boldmath{$\alpha_{1}$}}  &  \multirow{3}{*}{\boldmath{$\alpha_{2}$}}  & \multirow{3}{*}{ \boldmath{$E_b$} \textbf{(keV)} }\\
 & & & \boldmath{${\rm GeV}^{-2}$} \textbf{units for} \boldmath{$d=6$} & \boldmath{${\rm GeV}^{-2}$} \textbf{units for} \boldmath{$d=6$} & & & &  \\
 & & & \boldmath{${\rm GeV}^{-4}$} \textbf{units for} \boldmath{$d=8$} & \boldmath{${\rm GeV}^{-4}$} \textbf{untis for} \boldmath{$d=8$} & & & &    \\
\midrule
GRB 080916C & $(146.6^{\circ},\;119.8^{\circ})$ & 6 & $-1.12^{+3.72}_{-2.16} \times 10^{-13}$ & $-3.97^{+13.19}_{-7.66} \times 10^{-13}$ & $1.52^{+0.99}_{-1.39}$ & $3.61^{+4.78}_{-2.76}$ & $-5.73^{+4.45}_{-3.91}$ & $208.23^{+1541.98}_{-166.24}$ \\
 & & 8 & $-1.20^{+3.94}_{-3.25} \times 10^{-8}$ & $-4.25^{+13.97}_{-11.52} \times 10^{-8}$ & $1.51^{+0.98}_{-1.39}$ & $3.51^{+4.47}_{-2.57}$ & $-5.49^{+4.21}_{-3.97}$ & $219.22^{+1280.32}_{-180.51}$ \\
GRB 081221 & $(114.5^{\circ},\; 15.8^{\circ})$ & 6 & $6.49^{+6.92}_{-5.58} \times 10^{-12}$ & $23.01^{+24.53}_{-19.78} \times 10^{-12}$ & $2.59^{+3.19}_{-2.32}$ & $3.78^{+4.58}_{-2.65}$ & $-4.83^{+3.87}_{-4.69}$ & $11346.87^{+3412.24}_{-1269.44}$ \\
 & & 8 & $ 5.19^{+5.30}_{-5.12} \times 10^{-6}$ & $18.40^{+18.79}_{-18.15} \times 10^{-6}$ & $1.52^{+3.69}_{-1.36}$ & $3.80^{+4.99}_{-2.85}$ & $-5.24^{+4.45}_{-4.40}$ & $827.32^{+3793.61}_{-771.93}$ \\
GRB 090328 & $(56.6^{\circ},\; 155.7^{\circ})$ & 6 & $7.46^{+6.76}_{-6.66} \times 10^{-12}$ & $26.45^{+23.96}_{-23.61} \times 10^{-12}$ & $2.94^{+2.87}_{-2.46}$ & $2.18^{+3.50}_{-1.39}$ & $-2.42^{+2.60}_{-4.65}$ & $2042.26^{+2794.21}_{-1807.09}$ \\
 & & 8 & $6.54^{+4.15}_{-4.71} \times 10^{-6}$ & $23.18^{+14.71}_{-16.70} \times 10^{-6}$ & $1.30^{+3.48}_{-1.04}$ & $3.22^{+4.89}_{-2.47}$ & $-3.93^{+4.33}_{-5.38}$ & $1815.82^{+2981.54}_{-1786.95}$ \\
GRB 090618 & $(11.6^{\circ},\; 294.0^{\circ})$ & 6 & $1.52^{+1.39}_{-1.27} \times 10^{-11}$ & $5.39^{+4.93}_{-4.50} \times 10^{-11}$ & $3.11^{+0.85}_{-2.25}$ & $1.15^{+1.31}_{-0.55}$ & $0.13^{+0.42}_{-1.00}$ & $38.07^{+23.61}_{-31.46}$ \\
 & & 8 & $7.14^{+7.06}_{-6.63} \times 10^{-6}$ & $25.31^{+25.03}_{-23.50} \times 10^{-6}$ & $3.24^{+0.73}_{-2.19}$ & $1.48^{+1.27}_{-0.78}$ & $-0.23^{+0.57}_{-0.91}$ & $40.13^{+23.58}_{-31.24}$ \\\bottomrule
\end{tabularx}
\end{adjustwidth}
\end{table}

\begin{table}[H]\ContinuedFloat
\caption{{\em Cont.}}
\begin{adjustwidth}{-\extralength}{0cm}
\tablesize{\scriptsize}
\begin{tabularx}{\fulllength}{lCm{0.5cm}m{2.5cm}m{2.5cm}CCCC}
\toprule
\multirow{3}{*}{\textbf{Name}} & \multirow{3}{*}{\boldmath{$(\theta,\;\phi)$}} & \multirow{3}{*}{\boldmath{$d$}} & \boldmath{$\sum_{jm} {}_{0}Y_{jm}(\theta,\;\phi) c_{(I)jm}^{(d)}$} & \boldmath{$c_{(I)00}^{(d)}$} & \multirow{3}{*}{\boldmath{$\tau$}}  &  \multirow{3}{*}{\boldmath{$\alpha_{1}$}}  &  \multirow{3}{*}{\boldmath{$\alpha_{2}$}}  & \multirow{3}{*}{ \boldmath{$E_b$} \textbf{(keV)} }\\
 & & & \boldmath{${\rm GeV}^{-2}$} \textbf{units for} \boldmath{$d=6$} & \boldmath{${\rm GeV}^{-2}$} \textbf{units for} \boldmath{$d=6$} & & & &  \\
 & & & \boldmath{${\rm GeV}^{-4}$} \textbf{units for} \boldmath{$d=8$} & \boldmath{${\rm GeV}^{-4}$} \textbf{untis for} \boldmath{$d=8$} & & & &    \\
\midrule
GRB 090926A & $(156.3^{\circ},\; 353.4^{\circ})$ & 6 & $4.90^{+6.48}_{-5.71} \times 10^{-13}$ & $17.37^{+22.97}_{-20.24} \times 10^{-13}$ & $2.34^{+1.53}_{-1.88}$ & $0.87^{+0.98}_{-0.42}$ & $-1.70^{+1.46}_{-3.83}$ & $2101.57^{+2717.75}_{-2061.80}$ \\
 & & 8 & $8.16^{+9.12}_{-8.93} \times 10^{-8}$ & $28.93^{+32.33}_{-31.66} \times 10^{-8}$ & $1.94^{+1.31}_{-1.55}$ & $0.96^{+0.80}_{-0.43}$ & $-1.95^{+1.54}_{-2.42}$ & $1777.59^{+2989.77}_{-1709.57}$ \\
GRB 091003A & $(126.6^{\circ},\; 251.5^{\circ})$ & 6 & $2.70^{+3.69}_{-2.27} \times 10^{-13}$ & $9.57^{+13.08}_{-8.05} \times 10^{-13}$ & $0.21^{+0.17}_{-0.18}$ & $4.56^{+4.89}_{-3.38}$ & $-6.74^{+5.05}_{-3.08}$ & $225.55^{+1784.70}_{-152.93}$ \\
 & & 8 & $1.14^{+1.26}_{-1.21} \times 10^{-7}$ & $4.04^{+4.47}_{-4.29} \times 10^{-7}$ & $0.21^{+0.17}_{-0.16}$ & $4.72^{+4.83}_{-3.54}$ & $-7.12^{+4.71}_{-2.73}$ & $200.75^{+378.46}_{-127.32}$ \\
GRB 100728A & $(105.3^{\circ},\; 88.8^{\circ})$ & 6 & $-0.73^{+10.34}_{-5.84} \times 10^{-12}$ & $-2.59^{+36.65}_{-20.70} \times 10^{-12}$ & $2.61^{+1.33}_{-2.11}$ & $4.75^{+4.26}_{-2.48}$ & $-8.02^{+6.50}_{-1.89}$ & $378.99^{+1125.68}_{-347.10}$ \\
 & & 8 & $0.40^{+9.56}_{-5.86} \times 10^{-6}$ & $1.42^{+33.89}_{-20.77} \times 10^{-6}$ & $2.76^{+1.19}_{-2.13}$ & $6.89^{+2.90}_{-3.54}$ & $-3.23^{+3.41}_{-3.40}$ & $37.79^{+14.54}_{-19.55}$ \\
GRB 120119A & $(99.8^{\circ},\; 120.0^{\circ})$ & 6 & $3.11^{+3.90}_{-3.29} \times 10^{-12}$ & $11.02^{+13.83}_{-11.66} \times 10^{-12}$ & $1.65^{+2.16}_{-1.57}$ & $1.57^{+4.05}_{-1.04}$ & $-3.33^{+2.91}_{-5.43}$ & $1064.22^{+3638.78}_{-1003.59}$ \\
 & & 8 & $1.30^{+2.05}_{-1.66} \times 10^{-6}$ & $4.61^{+7.27}_{-5.88} \times 10^{-6}$ & $1.22^{+2.00}_{-1.15}$ & $2.17^{+4.71}_{-1.71}$ & $-4.15^{+3.71}_{-4.90}$ & $486.17^{+3803.39}_{-443.45}$ \\
GRB 130427A & $(62.3^{\circ},\; 173.1^{\circ})$ & 6 & $4.69^{+5.71}_{-6.08} \times 10^{-14}$ & $16.63^{+20.24}_{-21.55} \times 10^{-14}$ & $0.82^{+0.92}_{-0.69}$ & $2.57^{+2.79}_{-1.22}$ & $-0.14^{+0.11}_{-0.25}$ & $20.31^{+30.36}_{-12.97}$ \\
 & & 8 & $8.37^{+10.03}_{-10.31} \times 10^{-10}$ & $29.67^{+35.56}_{-36.55} \times 10^{-10}$ & $0.78^{+0.95}_{-0.73}$ & $2.63^{+3.95}_{-1.25}$ & $-0.18^{+0.10}_{-0.20}$ & $19.44^{+30.43}_{-13.81}$ \\
GRB 130518A & $(42.5^{\circ},\; 355.7^{\circ})$ & 6 & $6.98^{+6.26}_{-5.83} \times 10^{-14}$ & $24.74^{+22.19}_{-20.67} \times 10^{-14}$ & $0.75^{+0.59}_{-0.63}$ & $0.87^{+0.65}_{-0.37}$ & $-2.24^{+1.84}_{-2.62}$ & $1661.40^{+3072.03}_{-1498.65}$ \\
 & & 8 & $4.39^{+3.32}_{-3.48} \times 10^{-9}$ & $ 15.56^{+11.77}_{-12.34} \times 10^{-9}$ & $0.64^{+0.61}_{-0.52}$ & $0.90^{+0.74}_{-0.41}$ & $-2.52^{+2.19}_{-2.30}$ & $1353.03^{+3211.94}_{-1210.71}$ \\
GRB 130925A & $(116.1^{\circ},\; 41.2^{\circ})$ & 6 & $2.89^{+12.37}_{-7.01} \times 10^{-11}$ & $ 10.24^{+43.85}_{-24.85} \times 10^{-11}$ & $3.59^{+0.40}_{-1.67}$ & $1.06^{+0.42}_{-0.21}$ & $-9.46^{+1.84}_{-0.52}$ & $231.66^{+151.90}_{-29.60}$ \\
 & & 8 & $1.22^{+10.64}_{-3.59} \times 10^{-4}$ & $4.32^{+37.72}_{-12.73} \times 10^{-4}$ & $3.63^{+0.36}_{-2.52}$ & $1.05^{+3.49}_{-0.21}$ & $-9.51^{+1.68}_{-0.47}$ & $226.42^{+637.92}_{-22.97}$ \\
GRB 131108A & $(80.3^{\circ},\; 156.5^{\circ})$ & 6 & $0.08^{+4.46}_{-2.91} \times 10^{-13}$ & $0.28^{+15.81}_{-10.32} \times 10^{-13}$ & $0.68^{+0.41}_{-0.59}$ & $3.94^{+3.78}_{-2.61}$ & $-6.00^{+4.66}_{-3.60}$ & $331.23^{+955.10}_{-277.03}$ \\
 & & 8 & $-3.20^{+11.17}_{-10.96} \times 10^{-8}$ & $-11.34^{+39.60}_{-38.85} \times 10^{-8}$ & $0.61^{+0.42}_{-0.58}$ & $3.79^{+4.05}_{-2.55}$ & $-6.51^{+4.23}_{-3.22}$ & $346.53^{+1276.27}_{-271.00}$ \\
GRB 131231A & $(91.6^{\circ},\; 10.6^{\circ})$ & 6 & $1.04^{+0.42}_{-0.44} \times 10^{-11}$ & $3.69^{+1.49}_{-1.56} \times 10^{-11}$ & $3.83^{+0.16}_{-1.43}$ & $1.19^{+0.99}_{-0.19}$ & $0.11^{+0.17}_{-0.86}$ & $32.04^{+2.56}_{-15.21}$ \\
 & & 8 & $5.02^{+1.65}_{-1.41} \times 10^{-6}$ & $ 17.80^{+5.85}_{-5.00} \times 10^{-6}$ & $3.65^{+0.34}_{-1.61}$ & $1.94^{+0.63}_{-0.61}$ & $-0.58^{+0.53}_{-0.47}$ & $29.89^{+4.71}_{-15.90}$ \\
GRB 140206A & $(23.2^{\circ},\; 145.3^{\circ})$ & 6 & $0.85^{+1.03}_{-0.77} \times 10^{-12}$ & $3.01^{+3.65}_{-2.73} \times 10^{-12}$ & $1.33^{+0.82}_{-1.16}$ & $1.95^{+1.27}_{-0.74}$ & $-3.06^{+2.31}_{-2.58}$ & $413.97^{+2178.15}_{-390.50}$ \\
 & & 8 & $0.57^{+0.58}_{-0.48} \times 10^{-6}$ & $2.02^{+2.06}_{-1.70} \times 10^{-6}$ & $1.20^{+0.68}_{-1.10}$ & $2.10^{+1.22}_{-0.81}$ & $-3.21^{+3.01}_{-2.86}$ & $321.23^{+2120.21}_{-302.84}$ \\
GRB 140508A & $(43.2^{\circ},\; 255.5^{\circ})$ & 6 & $1.70^{+1.59}_{-1.52} \times 10^{-12}$ & $ 6.03^{+5.64}_{-5.39} \times 10^{-12}$ & $1.39^{+2.33}_{-1.26}$ & $0.50^{+1.11}_{-0.26}$ & $-1.28^{+1.62}_{-7.49}$ & $2756.68^{+2133.35}_{-2504.72}$ \\
 & & 8 & $0.73^{+0.73}_{-0.71} \times 10^{-6}$ & $ 2.59^{+2.59}_{-2.52} \times 10^{-6}$ & $0.74^{+1.71}_{-0.68}$ & $0.45^{+1.77}_{-0.29}$ & $-1.51^{+1.75}_{-6.88}$ & $2388.69^{+2482.19}_{-2274.19}$ \\
GRB 141028A & $(90.2^{\circ},\; 322.6^{\circ})$ & 6 & $2.13^{+4.42}_{-2.76} \times 10^{-12}$ & $7.55^{+15.67}_{-9.78} \times 10^{-12}$ & $2.70^{+3.11}_{-2.35}$ & $3.64^{+4.71}_{-2.90}$ & $-4.49^{+4.50}_{-5.02}$ & $1703.81^{+3102.34}_{-1578.31}$ \\
 & & 8 & $1.04^{+1.63}_{-1.53} \times 10^{-6}$ &  $3.69^{+5.78}_{-5.42} \times 10^{-6}$ & $1.40^{+2.43}_{-1.26}$ & $1.58^{+6.02}_{-1.15}$ & $-3.41^{+3.54}_{-5.91}$ & $1269.89^{+3487.24}_{-1183.83}$ \\
GRB 150314A & $(26.2^{\circ},\; 126.7^{\circ})$ & 6 & $1.25^{+2.38}_{-2.68} \times 10^{-13}$ & $4.43^{+8.44}_{-9.50} \times 10^{-13}$ & $2.13^{+3.58}_{-1.90}$ & $2.92^{+5.40}_{-2.36}$ & $-3.60^{+3.86}_{-5.85}$ & $1852.58^{+2959.22}_{-1735.94}$ \\
 & & 8 & $0.44^{+3.81}_{-5.44} \times 10^{-7}$ & $1.56^{+13.51}_{-19.28} \times 10^{-7}$ & $0.78^{+0.62}_{-0.61}$ & $0.53^{+0.47}_{-0.15}$ & $-1.97^{+1.58}_{-3.02}$ & $1877.20^{+2926.92}_{-1766.16}$ \\
GRB 150403A & $(152.7^{\circ},\; 311.5^{\circ})$ & 6 & $7.14^{+8.92}_{-8.00} \times 10^{-13}$ & $25.31^{+31.62}_{-28.36} \times 10^{-13}$ & $2.22^{+1.68}_{-1.92}$ & $0.59^{+1.10}_{-0.27}$ & $-1.64^{+1.94}_{-7.24}$ & $2491.52^{+2374.40}_{-2182.70}$ \\
 & & 8 & $1.94^{+1.76}_{-1.83} \times 10^{-7}$ & $ 6.88^{+6.24}_{-6.49} \times 10^{-7}$ & $1.96^{+1.87}_{-1.65}$ & $0.50^{+1.02}_{-0.22}$ & $-1.27^{+1.56}_{-7.51}$ & $2676.79^{+2202.30}_{-2345.65}$ \\
GRB 150514A & $(150.9^{\circ},\; 74.8^{\circ})$ & 6 & $1.61^{+2.41}_{-1.56} \times 10^{-11}$ & $5.71^{+8.54}_{-5.53} \times 10^{-11}$ & $1.79^{+2.08}_{-1.65}$ & $0.63^{+1.16}_{-0.37}$ & $-1.80^{+2.15}_{-7.32}$ & $2456.03^{+2411.40}_{-2284.19}$ \\
 & & 8 & $9.66^{+9.59}_{-8.99} \times 10^{-5}$ & $34.24^{+34.00}_{-31.87} \times 10^{-5}$ & $1.34^{+2.41}_{-1.23}$ & $0.51^{+0.95}_{-0.30}$ & $-1.41^{+1.76}_{-7.49}$ & $2683.10^{+2200.86}_{-2465.27}$ \\
GRB 150821A & $(147.9^{\circ},\; 341.9^{\circ})$ & 6 & $0.68^{+1.15}_{-0.61} \times 10^{-10}$ & $2.41^{+4.08}_{-2.16} \times 10^{-10}$ & $2.57^{+1.37}_{-2.20}$ & $1.56^{+3.22}_{-0.72}$ & $-4.35^{+5.28}_{-4.84}$ & $1065.04^{+3724.78}_{-1051.64}$ \\
 & & 8 & $1.69^{+1.52}_{-1.39} \times 10^{-4}$ & $5.99^{+5.39}_{-4.91} \times 10^{-4}$ & $3.08^{+0.88}_{-2.34}$ & $2.35^{+3.84}_{-1.52}$ & $-0.60^{+1.24}_{-3.14}$ & $43.75^{+22.32}_{-31.81}$ \\
GRB 160509A & $(14.0^{\circ},\; 310.1^{\circ})$ & 6 & $5.57^{+11.89}_{-10.90} \times 10^{-13}$ & $19.75^{+42.15}_{-38.64} \times 10^{-13}$ & $1.53^{+2.26}_{-1.31}$ & $0.96^{+1.51}_{-0.50}$ & $-1.77^{+2.01}_{-4.02}$ & $1933.51^{+2888.12}_{-1893.44}$ \\
 & & 8 & $0.93^{+3.49}_{-3.20} \times 10^{-7}$ & $3.30^{+12.37}_{-11.34} \times 10^{-7}$ & $1.18^{+1.75}_{-1.02}$ & $1.10^{+1.31}_{-0.64}$ & $-2.19^{+2.10}_{-3.36}$ & $1673.00^{+3087.91}_{-1572.41}$ \\
GRB 160625B & $(83.1^{\circ},\; 308.6^{\circ})$ & 6 & $3.63^{+0.57}_{-0.59} \times 10^{-15}$ & $12.87^{+2.02}_{-2.09} \times 10^{-15}$ & $0.62^{+0.18}_{-0.40}$ & $2.19^{+5.19}_{-1.47}$ & $0.33^{+0.03}_{-0.06}$ & $64.02^{+39.56}_{-27.87}$\\
 & & 8 & $1.98^{+0.29}_{-0.30} \times 10^{-12}$ & $ 7.02^{+1.03}_{1.06} \times 10^{-12}$ & $0.62^{+0.27}_{-0.43}$ & $2.15^{+4.14}_{-0.96}$ & $0.28^{+0.02}_{-0.03}$ & $66.47^{+38.56}_{-32.10}$\\
GRB 171010A & $(100.5^{\circ},\; 66.6^{\circ})$ & 6 & $3.45^{+4.11}_{-12.45} \times 10^{-11}$ & $12.23^{+14.57}_{-44.13} \times 10^{-11}$ & $3.01^{+0.97}_{-2.24}$ & $1.99^{+1.38}_{-0.88}$ & $-0.17^{+0.69}_{-5.96}$ & $23.77^{+147.73}_{-17.42}$ \\
 & & 8 & $2.43^{+2.58}_{-2.44} \times 10^{-5}$ & $8.61^{+9.15}_{-8.65} \times 10^{-5}$ & $3.04^{+0.92}_{-2.30}$ & $2.39^{+1.19}_{-0.89}$ & $-0.54^{+0.71}_{-0.76}$ & $24.32^{+10.24}_{-17.56}$ \\
GRB 180703A & $(157.1^{\circ},\; 6.5^{\circ})$ & 6 & $2.44^{+7.37}_{-3.32} \times 10^{-12}$ & $ 8.65^{+26.13}_{-11.77} \times 10^{-12}$ & $0.84^{+0.40}_{-0.65}$ & $5.84^{+3.80}_{-3.46}$ & $-6.94^{+6.00}_{-2.91}$ & $168.46^{+1201.15}_{-110.56}$ \\
 & & 8 & $3.20^{+7.01}_{-5.63} \times 10^{-6}$ & $11.34^{+24.85}_{-19.96} \times 10^{-6}$ & $0.80^{+0.23}_{-0.54}$ & $6.16^{+3.55}_{-3.73}$ & $-7.05^{+6.10}_{-2.81}$ & $143.70^{+257.42}_{-85.73}$ \\
GRB 180720B & $(93.0^{\circ},\; 0.59^{\circ})$ & 6 & $ -0.03^{+0.70}_{-0.65} \times 10^{-14}$ & $-0.11^{+2.48}_{-2.30} \times 10^{-14}$ & $2.92^{+0.43}_{-1.33}$ & $1.04^{+0.46}_{-0.38}$ & $-0.80^{+0.66}_{-0.63}$ & $2800.03^{+2092.83}_{-2465.91}$ \\
 & & 8 & $-0.13^{+0.93}_{-0.90} \times 10^{-11}$ & $-0.46^{+3.30}_{-3.19} \times 10^{-11}$ & $2.95^{+0.33}_{-2.05}$ & $1.06^{+0.65}_{-0.39}$ & $-0.81^{+0.63}_{-0.57}$ & $2785.79^{+2106.16}_{-2664.13}$ \\
GRB 190114C & $(116.9^{\circ},\; 54.5^{\circ})$ & 6 & $5.84^{+2.44}_{-2.34} \times 10^{-14}$ & $20.70^{+8.65}_{-8.30} \times 10^{-14}$ & $0.70^{+0.66}_{-0.61}$ & $1.54^{+0.77}_{-0.57}$ & $-3.04^{+1.61}_{-1.89}$ & $877.77^{+3537.70}_{-810.57}$ \\
 & & 8 & $0.97^{+0.56}_{-0.55} \times 10^{-9}$ & $ 3.44^{+1.99}_{-1.95} \times 10^{-9}$ & $0.34^{+0.94}_{-0.29}$ & $1.48^{+0.70}_{-0.63}$ & $-3.88^{+1.68}_{-1.77}$ & $1560.07^{+3096.27}_{-1352.64}$ \\
GRB 200613A & $(44.2^{\circ},\; 153.0^{\circ})$ & 6 & $0.59^{+0.84}_{-0.72} \times 10^{-11}$ & $2.09^{+2.98}_{-2.55} \times 10^{-11}$ & $2.02^{+1.86}_{-1.69}$ & $3.78^{+4.73}_{-3.16}$ & $-4.24^{+4.12}_{-5.24}$ & $1841.73^{+2983.38}_{-1748.44}$ \\
 & & 8 & $1.21^{+1.61}_{-1.64} \times 10^{-5}$ & $4.29^{+5.71}_{-5.81} \times 10^{-5}$ & $1.65^{+2.18}_{-1.38}$ & $3.26^{+5.14}_{-2.75}$ & $-3.68^{+3.67}_{-5.75}$ & $1835.19^{+2970.97}_{-1746.47}$ \\
GRB 200829A & $(17.6^{\circ},\; 251.1^{\circ})$ & 6 & $3.25^{+0.91}_{-1.06} \times 10^{-14}$ & $11.52^{+3.23}_{-3.76} \times 10^{-14}$ & $0.70^{+0.26}_{-0.39}$ & $0.79^{+0.62}_{-0.29}$ & $-1.61^{+1.22}_{-1.34}$ & $2446.19^{+2432.79}_{-2406.95}$ \\
 & & 8 & $4.40^{+2.86}_{-1.54} \times 10^{-10}$ & $15.60^{+10.14}_{-5.46} \times 10^{-10}$ & $0.36^{+0.45}_{-0.21}$ & $0.56^{+1.61}_{-0.11}$ & $-2.76^{+1.76}_{-1.74}$ & $3033.55^{+1850.99}_{-2989.83}$ \\
GRB 201216C & $(73.5^{\circ},\; 16.4^{\circ})$ & 6 & $0.92^{+2.53}_{-2.53} \times 10^{-12}$ & $3.26^{+8.97}_{-8.97} \times 10^{-12}$ & $2.03^{+1.79}_{-1.77}$ & $0.75^{+0.80}_{-0.39}$ & $-1.92^{+2.06}_{-4.95}$ & $1778.23^{+3002.73}_{-1764.69}$ \\
 & & 8 & $0.08^{+1.00}_{-1.11} \times 10^{-6}$ & $0.28^{+3.54}_{-3.93} \times 10^{-6}$ & $1.64^{+1.55}_{-1.44}$ & $0.82^{+0.88}_{-0.41}$ & $-2.28^{+1.78}_{-3.35}$ & $1479.39^{+3212.11}_{-1331.33}$ \\
GRB 210204A & $(80.3^{\circ},\; 109.1^{\circ})$ & 6 & $0.17^{+11.89}_{-8.90} \times 10^{-12}$ & $0.60^{+42.04}_{-31.55} \times 10^{-12}$ & $2.37^{+1.60}_{-2.16}$ & $1.27^{+2.03}_{-0.48}$ & $-3.91^{+3.87}_{-4.64}$ & $308.84^{+2115.32}_{-243.46}$ \\
 & & 8 & $0.30^{+1.40}_{-1.30} \times 10^{-5}$ & $1.06^{+4.96}_{-4.61} \times 10^{-5}$ & $2.77^{+1.21}_{-2.54}$ & $1.30^{+2.31}_{-0.52}$ & $-3.57^{+3.47}_{-4.14}$ & $274.49^{+1155.77}_{-209.90}$ \\
GRB 210610B & $(75.6^{\circ},\; 243.9^{\circ})$ & 6 & $-0.77^{+10.75}_{-8.37} \times 10^{-12}$ & $-2.73^{+38.11}_{-29.67} \times 10^{-12}$ & $4.30^{+3.82}_{-3.87}$ & $3.76^{+3.68}_{-2.68}$ & $-5.72^{+3.87}_{-3.46}$ & $321.33^{+1719.51}_{-287.08}$ \\
 & & 8 & $-2.60^{+11.34}_{-11.61} \times 10^{-6}$ & $-9.22^{+40.20}_{-41.16} \times 10^{-6}$ & $3.94^{+4.10}_{-3.82}$ & $3.88^{+3.35}_{-2.83}$ & $-5.99^{+4.09}_{-3.46}$ & $331.28^{+2465.18}_{-297.21}$ \\
GRB 210619B & $(56.1^{\circ},\;319.7^{\circ})$ & 6 & $1.20^{+0.45}_{-0.46} \times 10^{-15}$ & $4.25^{+1.60}_{-1.63} \times 10^{-15}$ & $0.23^{+0.22}_{-0.18}$ & $2.02^{+2.47}_{-0.91}$ & $0.24^{+0.03}_{-0.04}$ & $36.71^{+45.06}_{-23.05}$ \\
 & & 8 & $1.52^{+0.52}_{-0.52} \times 10^{-12}$ & $5.38^{+1.84}_{-1.84} \times 10^{-12}$ & $0.23^{+0.23}_{-0.18}$ & $2.00^{+2.33}_{-0.90}$ & $0.22^{+0.03}_{-0.04}$ & $36.84^{+48.26}_{-22.81}$ \\
\bottomrule
\end{tabularx}
\end{adjustwidth}
\end{table}

\vspace{-20pt}
{\captionof*{figure}{parameters 
($\tau$, $\alpha_{1}$, $\alpha_{2}$, $\beta$, and $E_{b}$) and vacuum coefficients with $d=6$ for GRB 130427A. 
Best-fit values are highlighted by vertical solid lines, whereas $\pm2\sigma$ deviations from the best-fit values are indicated by vertical dashed lines. Lower panel: same but for the analysis with $d=8$ coefficients.}}

\vspace{6pt}

\section{Discussion}
\label{sec:Discussion}
{Estimating from the 95\% confidence levels in Table~\ref{tab2}, most of our constraints on 
$\sum_{jm} {}_{0}Y_{jm}(\hat{\textbf{\emph{n}}}) c_{(I)jm}^{(d)}$ are consistent with zero within 
the $3\sigma$ confidence level, implying that there is no convincing evidence for LIV.
However, the results from GRB 131231A, GRB 160625B, GRB 190114C, GRB 200829A, and GRB 210619B are 
incompatible with zero at the $4.7\sigma$, $12.3\sigma$, $5.0\sigma$, $6.1\sigma$, and $5.2\sigma$, 
respectively. Given the fact that there is a significant deviation from the null hypothesis of no LIV in 5 
out of 32 GRBs, here, we examine the statistical significance of the evidence for a non-zero LIV 
spectral lag. 
We fit the data to an additional hypothesis that the time lags are only due to 
the intrinsic astrophysical mechanisms given by Equation~(\ref{eq:tint}). Once we obtain the 
best-fit parameters, we then proceed to carry out model comparison using the Akaike Information 
Criterion (AIC) by treating the case of only intrinsic astrophysical emission as the null hypothesis.
The AIC score of each fitted model is given as ${\rm AIC}=\chi^{2}+2f$, where $f$ is the number of model-free
parameters \citep{1974ITAC...19..716A,2007MNRAS.377L..74L}. With ${\rm AIC}_i$ characterizing model $\mathcal{M}_i$,
the un-normalized confidence that this model is true is the Akaike weight $\exp(-{\rm AIC}_{i}/2)$. 
The relative probability of $\mathcal{M}_i$ being the correct model in a one-to-one comparison is then 
\begin{equation}
P(\mathcal{M}_i)=\frac{\exp(-{\rm AIC}_i/2)}{\exp(-{\rm AIC_1}/2)+\exp(-{\rm AIC_2}/2)}\;.
\end{equation}
It is easy to see that the model with less AIC score is the one more preferred by this criterion. 
According to the AIC model selection criterion, we find that compared with the LIV ($d=6$) model, 
the no-LIV model can be discarded as having a probability of being correct of only $\sim 10^{-6}$ 
for GRB 131231A, $\sim 10^{-34}$ for GRB 160625B, $\sim 10^{-6}$ for GRB 190114C, $\sim 10^{-8}$ 
for GRB 200829A, and $\sim 10^{-6}$ for GRB 210619B. 
These results suggest that the evidence against
the null hypothesis is very strong. However, the deviation of spectral-lag behavior from the SBPL model, 
even if present in these five GRBs, cannot be due to Lorentz-violating effects, as this would contradict 
with previous upper limits and must therefore be of intrinsic astrophysical origin. Since the SBPL model
does not correctly describe the intrinsic lag behavior of these five GRBs, their resulting constraints 
on LIV can only be considered as artificial upper limits. 
In other words, if a more accurate model for 
the intrinsic lag behavior is adopted for these five GRBs, more stringent limits on LIV could be achieved.} 

It is worth emphasizing that
most of the previous works were carried out by concentrating on the rough time lag of a single highest-energy photon
and neglecting the source-intrinsic time lag. Performing a search for Lorentz-violating effects using the true
multi-photon spectral-lag data and considering the intrinsic time-lag problem, as performed in this work, is therefore
crucial. 
Although our method relies on a particular model of the GRB lag transition, our resulting constraints 
can still be deemed as comparatively robust.

In our analysis, we adopt the Hubble constant $H_{0}=67.36$ km $\rm s^{-1}$ $\rm Mpc^{-1}$ inferred from $Planck$ 
cosmic microwave background observations under $\Lambda$CDM \citep{2020A&A...641A...6P}. However, this value is 
in $5\sigma$ tension with that measured from local distance ladders ($H_{0}=73.04$ km $\rm s^{-1}$ $\rm Mpc^{-1}$)
\citep{2022ApJ...934L...7R}. Given the Hubble tension, we next consider whether the $H_{0}$ prior affects our constraints 
on the SME coefficients. We also perform a parallel analysis of the lag-energy data of GRB 130427A using a different $H_{0}$
prior. When $H_{0}$ varies from $67.36$ to $73.04$ km $\rm s^{-1}$ $\rm Mpc^{-1}$, we find that the limit on the combination 
$\sum_{jm} {}_{0}Y_{jm}(62.3^{\circ},\; 173.1^{\circ}) c_{(I)jm}^{(6)}$ of SME coefficients varies from 
$4.69^{+5.71}_{-6.08} \times 10^{-14}$ ${\rm GeV}^{-2}$ to $4.62^{+5.84}_{-7.13} \times 10^{-14}$ ${\rm GeV}^{-2}$. 
It is obvious that the choice of a different $H_{0}$ value has only a minimal influence on the results.

Liu et al. \cite{2022arXiv220209999L} derived limits on isotropic linear and quadratic leading-order
Lorentz-violating vacuum dispersion using the same 32 GRBs. Since the linear and quadratic LIV parameters 
in the vacuum isotropic model are global and direction-independent, they were able to obtain the statistical distributions of the LIV parameters upon fitting the entire sample of 32 GRBs (see Figure 3 of \cite{2022arXiv220209999L}).
In this work, however, we consider anisotropic Lorentz-violating vacuum dispersion, along with direction-dependent
effects. For each GRB, we derive a limit on one direction-specific combination 
$\sum_{jm} {}_{0}Y_{jm}(\hat{\textbf{\emph{n}}}) c_{(I)jm}^{(d)}$ of Lorentz-violating coefficients. 
Thus, it is not feasible to conduct a global fit on the LIV parameters by taking into account all GRBs 
(as Liu et al. \cite{2022arXiv220209999L} carried out in their treatment) in the vacuum anisotropic model. 

\section{Summary}
\label{sec:summary}

Lorentz-violating effects on the vacuum propagation of electromagnetic waves, allowing for arbitrary mass dimension $d$,
are well described in the SME effective field theory. Operators of odd dimension $d$ lead to both an energy-dependent 
vacuum dispersion and vacuum birefringence, whereas for each even $d$ there is a subset of $(d-1)^{2}$ nonbirefringent
Lorentz-violating operators leading to an anisotropic photon dispersion. Lorentz-violating terms with even $d$ are
characterized by a set of $(d-1)^{2}$ real coefficients, which can be constrained through astrophysical time-of-flight
measurements from $(d-1)^{2}$ or more directions in the sky. In this work, we focus on measuring coefficients
controlling nonbirefringent vacuum dispersion with $d=6$ and $8$.

We perform a search for Lorentz-violating photon dispersion from the muti-photon spectral-lag data of 32 $Fermi$ GRBs, 
all of whom have well-defined transitions from positive to negative spectral lags. The spectral-lag transitions can 
help to distinguish the possible time delay induced by Lorentz-violating effects from any source-intrinsic time delay
in different energy bands. Moreover, unlike most of the previous works that rely on using the rough time lag of a single 
highest-energy photon, these 32 GRBs have high photon statistics, allowing the use of the true time lags of broad 
light curves in multi-photon bands of different energies. By fitting the spectral-lag data of 32 GRBs,
we obtain robust constraints on coefficients for Lorentz violation. The results are listed in Table~\ref{tab2}.

A compilation of existing astrophysical limits on Lorentz-violating coefficients obtained through the photon dispersion
method can be found in ref. \cite{2011RvMP...83...11K}. For the case of $d=6$, our constraints are not competitive with
existing bounds but can be deemed as comparatively robust. For $d=8$, only a few bounds were obtained on the 
49 coefficients for nonbirefringent dispersion. 
Our new constraints have the promise to complement existing SME constraints.

\vspace{6pt} 



\authorcontributions{Conceptualization, Jun-Jie Wei and Xue-Feng Wu; Formal analysis, Jin-Nan Wei; Investigation, Jin-Nan Wei; Methodology, Zi-Ke Liu and Bin-Bin Zhang; Project administration, Jun-Jie Wei; Software, Zi-Ke Liu and Bin-Bin Zhang; Supervision, Jun-Jie Wei and Xue-Feng Wu; Validation, Zi-Ke Liu and Bin-Bin Zhang; Writing – original draft, Jin-Nan Wei and Jun-Jie Wei; Writing – review \& editing, Bin-Bin Zhang and Xue-Feng Wu.}

\funding{This research was funded by the National Key Research and Development Programs of China (2018YFA0404204), 
the National Natural Science Foundation of China (grant Nos.~11725314, 12041306, 11833003, U2038105, and 12121003), 
the Key Research Program of Frontier Sciences (grant No. ZDBS-LY-7014) of Chinese Academy of Sciences, 
the Major Science and Technology Project of Qinghai Province (2019-ZJ-A10), the Natural Science Foundation of 
Jiangsu Province (grant No. BK20221562), the China Manned Space Project (CMS-CSST-2021-B11), the Guangxi Key 
Laboratory for Relativistic Astrophysics, and the Program for Innovative Talents, Entrepreneur in Jiangsu.}



\dataavailability{The data underlying this article are available in the article and Liu et~al.~\cite{2022arXiv220209999L}.}

\acknowledgments{We are grateful to the anonymous referees for helpful comments.
J.-J.W. would like to thank the participants of the Ninth Meeting on CPT and Lorentz Symmetry 
(Bloomington, Indiana, 17--26 May 2022) for helpful discussions that led to the creation of 
this article. We also acknowledge the use of public data from the Fermi Science Support Center (FSSC).}

\conflictsofinterest{The authors declare no conflict of interest.} 


\clearpage

\begin{adjustwidth}{-\extralength}{0cm}

\reftitle{References}



\end{adjustwidth}
\end{document}